\documentclass[sigconf]{acmart}
\usepackage{tabularx}
\usepackage{booktabs} %
\usepackage{array} %
\usepackage{colortbl} %
\usepackage{graphicx}
\usepackage{tabularray}
\usepackage{soul}

\newcommand{\leftcell}[2][l]{%
  \begin{tabular}[#1]{@{}l@{}}#2\end{tabular}}

\usepackage{xspace}
\usepackage{subcaption}

\newcommand{\sysname}{ImaginateAR}
\newcommand{\rev}[1]{\textcolor{black}{#1}}

\AtBeginDocument{%
  }

\makeatletter
\def\@ACM@copyright@check@cc{}
\makeatother

\copyrightyear{2025}
\acmYear{2025}
\setcopyright{cc}
\setcctype{by}
\acmConference[UIST '25]{The 38th Annual ACM Symposium on User Interface Software and Technology}{September 28-October 1, 2025}{Busan, Republic of Korea}
\acmBooktitle{The 38th Annual ACM Symposium on User Interface Software and Technology (UIST '25), September 28-October 1, 2025, Busan, Republic of Korea}\acmDOI{10.1145/3746059.3747635}
\acmISBN{979-8-4007-2037-6/2025/09}

\begin{document}

\title{\sysname{}: AI-Assisted In-Situ Authoring in Augmented Reality}

\author{Jaewook Lee}
\authornote{Both authors contributed equally to this research.}
\affiliation{%
  \institution{University of Washington}
  \city{Seattle}
  \state{Washington}
  \country{USA}
}

\author{Filippo Aleotti}
\authornotemark[1]
\affiliation{%
  \institution{Niantic Spatial, Inc.}
  \city{London}
  \country{United Kingdom}
}

\author{Diego Mazala}
\affiliation{%
  \institution{Niantic Spatial, Inc.}
  \city{London}
  \country{United Kingdom}
}

\author{Guillermo Garcia-Hernando}
\affiliation{%
  \institution{Niantic Spatial, Inc.}
  \city{London}
  \country{United Kingdom}
}

\author{Sara Vicente}
\affiliation{%
  \institution{Niantic Spatial, Inc.}
  \city{London}
  \country{United Kingdom}
}

\author{Oliver James Johnston}
\affiliation{%
  \institution{Niantic Spatial, Inc.}
  \city{London}
  \country{United Kingdom}
}

\author{Isabel Kraus-Liang}
\affiliation{%
  \institution{Niantic Spatial, Inc.}
  \city{London}
  \country{United Kingdom}
}

\author{Jakub Powierza}
\affiliation{%
  \institution{Niantic Spatial, Inc.}
  \city{London}
  \country{United Kingdom}
}

\author{Donghoon Shin}
\affiliation{%
  \institution{University of Washington}
  \city{Seattle}
  \state{Washington}
  \country{USA}
}

\author{Jon E. Froehlich}
\affiliation{%
  \institution{University of Washington}
  \city{Seattle}
  \state{Washington}
  \country{USA}
}

\author{Gabriel Brostow}
\affiliation{%
  \institution{University College London \\ Niantic Spatial, Inc.}
  \city{London}
  \country{United Kingdom}
}

\author{Jessica Van Brummelen}
\affiliation{%
  \institution{Niantic Spatial, Inc.}
  \city{London}
  \country{United Kingdom}
}

\renewcommand{\shortauthors}{Lee and Aleotti, et al.}

\begin{abstract}
While augmented reality (AR) enables new ways to play, tell stories, and explore ideas rooted in the physical world, authoring personalized AR content remains difficult for non-experts, often requiring professional tools and time. Prior systems have explored AI-driven XR design but typically rely on manually defined VR environments and fixed asset libraries, limiting creative flexibility and real-world relevance. \rev{We introduce \textit{\sysname{}}, the first mobile tool for AI-assisted AR authoring to combine offline scene understanding, fast 3D asset generation, and LLMs---enabling users to create outdoor scenes through natural language interaction. For example, saying ``\textit{a dragon enjoying a campfire}'' (P7) prompts the system to generate and arrange relevant assets, which can then be refined manually.} Our technical evaluation shows that \rev{our custom pipelines} produce more accurate outdoor scene graphs and generate 3D meshes faster than prior methods. A three-part user study (N=20) revealed preferred roles for AI, how users create in free-form use, and design implications for future AR authoring tools. \rev{\sysname{} takes a step toward empowering anyone to create AR experiences anywhere---simply by speaking their imagination.}
\end{abstract}

\begin{teaserfigure}
  \includegraphics[width=\textwidth]{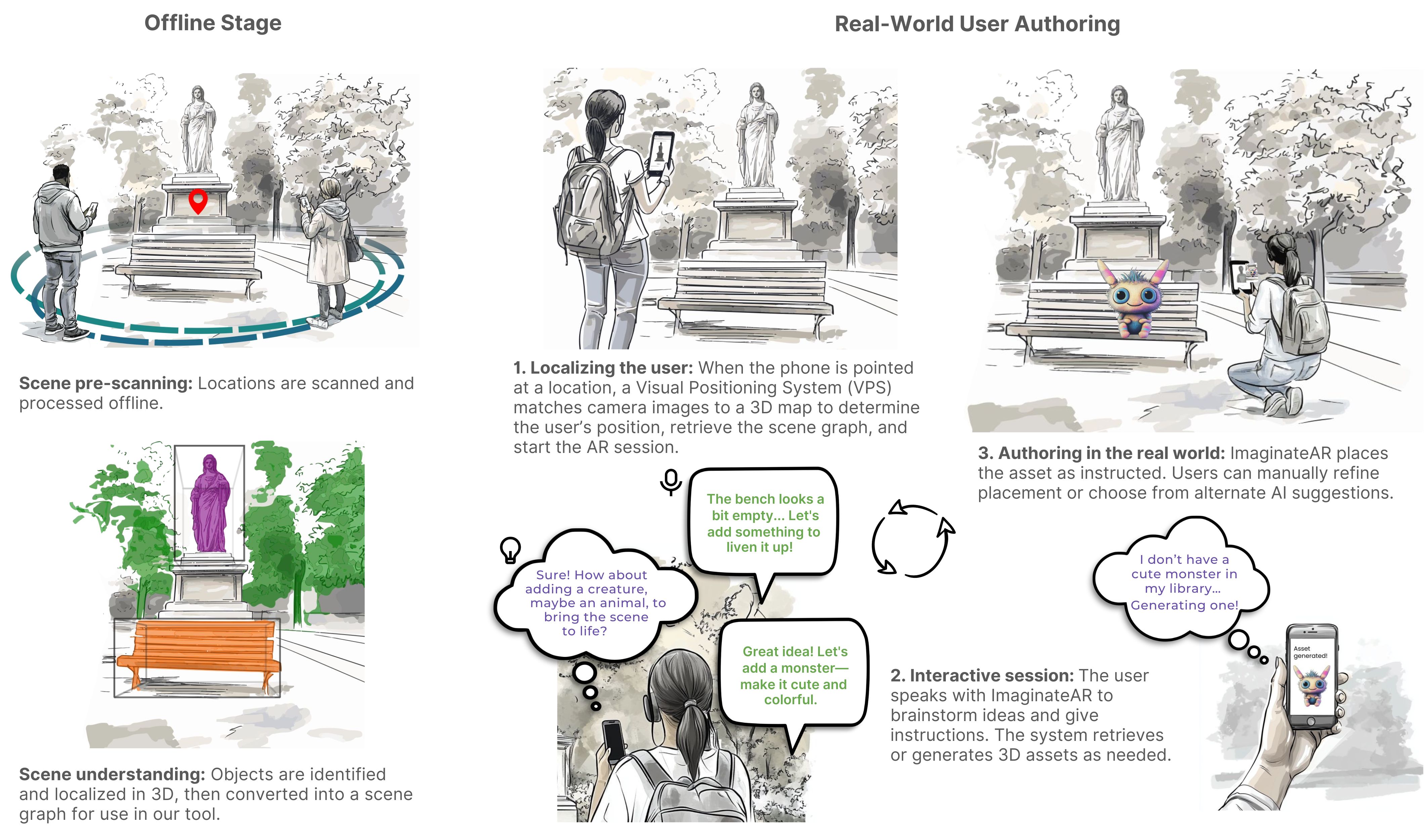}
  \caption{\sysname{} enables non-expert users to author personalized AR experiences \rev{through natural language interaction.} After a location is pre-scanned and processed by our scene understanding pipeline (left), users can brainstorm, generate, \rev{and place} virtual content on-site with AI assistance (right), and make manual adjustments as needed.
  }
  \label{fig:teaser}
\end{teaserfigure}

\maketitle

\section{Introduction}
Augmented Reality (AR) can transform everyday spaces into interactive canvases, blending digital content with the physical world. Today, AR is used not just for entertainment, but also to bring people together through games like \textit{Pokémon GO}~\cite{pokemonGO}, support location-based education~\cite{weAreStillHere-jauregui}, and amplify social causes through public art and storytelling~\cite{withOrWithoutPermissionARForSocialJustice-silva}. Yet, most AR content is created using professional tools like \textit{Unity}~\cite{unity}, \textit{Blender}~\cite{blender}, and \textit{Lens Studio}~\cite{lensStudio}, requiring specialized skills and limiting \textit{who} can create and \textit{what} is possible. While this enables highly polished experiences, it leaves everyday users without a way to easily and creatively customize their surroundings with AR. Imagine if everyone had the power to create their own AR worlds---teachers could build interactive history lessons in a schoolyard, artists could install digital murals on city walls, and friends could fill a beach with dancing penguins.

Although some consumer AR applications like \textit{Adobe Aero}~\cite{adobeAero}, \textit{IKEA Place}~\cite{ikeaPlace}, and \textit{LEGO AR Studio}~\cite{legoStudio} allow users to create AR content, they rely on predefined assets and manual placement, limiting creative flexibility and expressivity. To address these limitations, recent research has explored generative AI for authoring in extended reality (XR). For instance, systems like \textit{SceneCraft}~\cite{sceneCraftAnLlmAgentForBlender-hu}, \textit{3D-GPT}~\cite{3dGptProcedural3DModelingWithLlms-sun}, \textit{Ostaad}~\cite{howPeoplePromptVRAuthoring-manesh}, \rev{\textit{VRCopilot}~\cite{Zhang2024}}, \textit{LLMR}~\cite{llmr-delatorre}, \textit{LLMER}~\cite{llmer-chen}, \rev{and \textit{Dreamcrafter}~\cite{Vachha2025}} integrate large language models (LLMs) for XR scene generation and editing via natural language interaction. \rev{While promising, these systems primarily target manually defined environments} and lack in-situ authoring, real-world scene understanding, and/or open-ended asset generation, hindering truly personalized AR creation. Furthermore, most scene understanding algorithms are trained on indoor data~\cite{takmaz2023openmask3d,chen2024clip,gu2024conceptgraphs,peng2023openscene,conceptfusion}---so even if prior XR systems sought to incorporate scene understanding, existing models are not readily applicable to outdoor use---despite outdoor AR applications having proven impactful~\cite{pokemonGO,weAreStillHere-jauregui,withOrWithoutPermissionARForSocialJustice-silva}.

\rev{Suppose anyone could build an AR scene simply by speaking to an AI. A child might turn their backyard into a medieval kingdom by saying, ``\textit{Place a pink castle here.}'' and ``\textit{Add a fire-breathing dragon on the fence!}'' An urban planner could preview a structure with, ``\textit{Place a five-story apartment building here.}'' and ``\textit{Make it twice as tall!}'' And anyone could build just for fun. This is our vision for AR authoring: enabling users to create immersive scenes grounded in the real world by describing what they imagine. In this paper, we take a step toward that vision by supporting AR authoring in a wide range of static outdoor environments. We introduce \textit{\sysname{}}, the first mobile tool for AI-assisted AR authoring that generates and arranges virtual assets from speech input, facilitating their seamless integration into the physical world. \sysname{} achieves this by pushing the boundaries of (1) outdoor scene understanding, (2) fast 3D asset generation, and (3) LLM-driven natural language interaction---each a significant challenge for fully adaptive AR. Together, these advances help bring generative scene authoring---previously confined to VR---into real-world AR.}

To address real-world scene understanding, we updated open-vocabulary 3D instance segmentation models---typically trained on indoor data and reliant on user-specified queries---to function autonomously outdoors. Specifically, we enhance \textit{OpenMask3D}~\cite{takmaz2023openmask3d} with \textit{GPT-4o}~\cite{GPT4o} for consistent, automatic outdoor semantic labeling and apply \textit{HDBSCAN}~\cite{mcinnes2017hdbscan} clustering to merge redundant object masks. This produces structured scene graphs composed of labeled 3D bounding boxes, enabling spatial reasoning in real-world contexts. To improve usability and ensure a more complete view of the environment, we perform scene understanding offline on pre-scanned environments and retrieve the relevant scene graph at runtime using a \textit{Visual Positioning System (VPS)}~\cite{VPS}, rather than requiring users to scan live. For dynamic 3D mesh generation---essential for creativity and personalization---we contribute a pipeline that encourages well-formed AR assets (\textit{i.e.}, complete, volumetric, properly oriented, and scaled), while running significantly faster than prior methods. Our approach expands user input with GPT-4o, synthesizes reference images using \textit{Dall-E 2}~\cite{dallEEditing-openAi}, segments foreground objects via \textit{DIS}~\cite{qin2022}, and lifts them into 3D using \textit{InstantMesh}~\cite{xu2024instantmesh}. Finally, a multi-agent LLM pipeline enables speech-driven interaction: a \textit{Brainstorming} agent suggests scene ideas, an \textit{Action Plan} agent determines spatial relationships, and an \textit{Assembly} agent updates the scene graph for coherent placement.

To evaluate \sysname{}, we conducted a technical assessment of our scene understanding and asset generation pipelines, along with a three-part user study in a public park with 20 participants. Our scene understanding pipeline outperformed the base OpenMask3D~\cite{takmaz2023openmask3d} model and ablated variants of our pipeline, while our asset generation pipeline achieved comparable quality to state-of-the-art methods but with a significantly faster, sub-minute runtime. As part of our technical evaluation, we also conducted a demonstration-based assessment across varied outdoor scenes, showing that \sysname{} functions reliably beyond the user study setting. In the user study, participants first explored three authoring modes---\textit{manual}, \textit{AI-assisted}, and \textit{AI-decided}---to evaluate trade-offs between control and automation during different stages of AR authoring. They then used \sysname{} to freely design their own AR experiences (Part 2), followed by a co-design session to brainstorm future features (Part 3). Participants enjoyed interacting with \sysname{}, asking it to ``\textit{Put a dancing T-Rex on the grass}'' (P1) or ``\textit{Make a helicopter hover over the shed}'' (P14). Across sessions, users preferred a hybrid approach---leveraging AI for rapid and creative scene generation while retaining manual control for fine-tuned customization. AI assistance accelerated ideation and spatial arrangement, but participants often opted for manual refinement to ensure their creative intent was more precisely reflected in the final scene. We conclude by discussing current limitations and future directions for AI-assisted AR authoring.

In summary, our contributions include: (1) \sysname{}, a novel AI-assisted AR authoring tool that integrates real-world scene understanding, generative AI, and LLM-based reasoning to streamline content creation; (2) technical innovations in outdoor scene understanding, fast 3D asset generation, and a multi-agent LLM pipeline for speech interaction; and (3) insights into how users engage with AI-assisted AR authoring---including their balance of automation and control, free-form use, and desired future features. 

\section{Related Work}
We situate our work at the intersection of HCI and computer vision (CV), drawing from research on AI-powered XR authoring, real-world 3D scene understanding, generative AI for content creation, and AI assistance in creative workflows.

\subsection{AI-Powered XR Authoring}
\label{sec:xr-authoring-rel-work}
Because it involves 3D modeling, programming, and spatial design, creating XR content is inherently challenging~\cite{creatingARAndVRReview-ashtari,xrToolsAndWhereTakingUsReview-nebeling}. To lower this barrier, commercial tools like \textit{Adobe Aero}~\cite{adobeAero}, \textit{Unity MARS}~\cite{unityMars}, and \textit{Torch}~\cite{torch} offer direct manipulation interfaces for placing virtual objects, enabling users to manually design scenes, albeit without AI-driven automation or generation. Research prototypes such as \textit{Pronto}~\cite{prontoRapidArVideoPrototyping-leiva}, \textit{Rapido}~\cite{rapidoPrototypingInteractiveAr-leiva}, and \textit{ARAnimator}~\cite{arAnimator-ye} simplify AR prototyping through sketches and demonstration-based input, though they primarily support 2D content. Other systems, such as \textit{SemanticAdapt}~\cite{semanticAdapt-cheng}, \textit{ARTiST}~\cite{artistAutomatedTextSimplification-wu}, and Lindlbauer \textit{et al.}~\cite{contextAwareAdaptationMR-lindlbauer}, automate content arrangement based on scene semantics but focus on adaptive user interfaces rather than open-ended scene creation. In our work, we explore how generative AI and real-world scene understanding can further lower authoring barriers, \rev{taking a step toward} enabling anyone to create any AR experience.

Several recent systems have also explored using AI to streamline XR authoring. For instance, \textit{SonifyAR}~\cite{sonifyar-su} generates context-aware sound effects in mobile AR by leveraging LLMs. Others, such as \textit{BlenderGPT}~\cite{BlenderGPT}, \textit{SceneCraft}~\cite{sceneCraftAnLlmAgentForBlender-hu}, and \textit{3D-GPT}~\cite{3dGptProcedural3DModelingWithLlms-sun}, enable users to generate 3D models via natural language, which can later be arranged into virtual scenes---but they lack fast, in-situ authoring, limiting on-site ideation and iteration. More comprehensive tools like \textit{Ostaad}~\cite{howPeoplePromptVRAuthoring-manesh}, \textit{DreamCodeVR}~\cite{dreamCodeVR-giunchi}, \rev{\textit{VRCopilot}~\cite{Zhang2024}, \textit{Dreamcrafter}~\cite{Vachha2025},} and \textit{LLMR}~\cite{llmr-delatorre} go further by allowing users to iteratively prompt LLMs to build up full XR scenes. While these systems demonstrate the potential of LLM-assisted XR content creation, they primarily target VR \rev{and/or} rely on predefined asset libraries, limiting expressivity, adaptability, and real-world interaction. Closest to our work, \textit{LLMER}~\cite{llmer-chen} extends LLMR to mixed reality, and Fang \textit{et al.}~\cite{enablingWaypointGeneration-fang} integrate scene graphs, LLMs, and AR to facilitate robot navigation programming. However, both systems rely on manually constructed scene representations rather than automated scene understanding models. Ultimately, no existing system fully supports in-situ, speech-driven AR authoring with real-world scene understanding and open-ended asset generation. Prior work has also largely overlooked outdoor AR authoring, despite its proven impact in fun, education, public art, and social connection~\cite{pokemonGO,weAreStillHere-jauregui,withOrWithoutPermissionARForSocialJustice-silva}.

Building on this foundational work, we explore how outdoor scene understanding, fast 3D mesh generation, and LLM-driven speech interaction \rev{can help bring AI-assisted scene authoring---once limited to VR---into real-world AR.}

\subsection{Real-World 3D Scene Understanding}
\label{sec:scene-understanding}
Understanding real-world environments is a fundamental challenge for AR and robotics applications~\cite{billinghurst2015survey, cadena2016past}. To seamlessly integrate virtual content into physical spaces, systems must capture both \textit{geometric} and \textit{semantic} properties of a scene. Typically, this is achieved in two steps: first, a 3D map of the environment is built using cameras~\cite{newcombe2010live,sayed2022simplerecon}, sometimes augmented with depth or IMU sensors~\cite{newcombe2011kinectfusion,dai2017bundlefusion}. Next, semantic labels are assigned through CV models trained on 3D datasets~\cite{engelmann2017exploring,dai2017scannet}, enabling object recognition~\cite{schult2023mask3d,takmaz2023openmask3d}. Beyond individual object detection, some systems structure this information into \textit{scene graphs}~\cite{armeni20193d,rosinol20203d,Wald2020scenegraph,wu2021scenegraphfusion,koch2024open3dsg}, where objects are nodes and relationships (\textit{e.g.,} ``\textit{a bench is next to a tree}'') form edges. This structured representation enables high-level reasoning for context-aware applications, including ours.

Recent advances in multimodal models, such as \textit{CLIP}~\cite{radford2021learning} and vision-language models (VLMs), have enabled open-world object detection~\cite{takmaz2023openmask3d,chen2024clip,gu2024conceptgraphs,peng2023openscene,conceptfusion}, allowing models to recognize objects beyond predefined labels. This is critical for real-world use, as environments vary widely---indoor spaces differ from outdoor settings, and even rural and suburban outdoor areas contain distinct objects. Recent efforts in open-vocabulary scene understanding have integrated 3D cues directly into LLMs~\cite{huang2023embodied,yang2024llm,ma2024llms}, enabling agents to perform grounding, question-answering, and captioning within 3D environments. 
While promising, most open-vocabulary segmentation models rely on query-based retrieval~\cite{takmaz2023openmask3d}, identifying scene objects via user prompts or predefined vocabularies. This poses challenges for generating scene graphs: user prompts introduce latency during live graph construction for AR authoring, while defining a single comprehensive vocabulary for arbitrary scenes---needed for offline computation---is difficult. Furthermore, prior work has largely focused on indoor spaces, where object categories are more constrained and fundamentally different from those outdoors.

As such, we explore how existing open-vocabulary 3D instance segmentation models could be updated for outdoor AR---enabling \sysname{} to generate structured scene graphs of diverse environments through an automatic, offline scene understanding pipeline.

\subsection{Generative AI for Content Creation}
\label{sec:content-generation} 
\sysname{} leverages generative AI for fast, open-ended 3D asset creation, allowing users to verbally generate objects on demand---supporting creative flexibility and expressivity. Traditionally, 3D models are crafted by experts using professional tools, a time-consuming process infeasible for everyday users. While generative models have significantly advanced in 2D content creation, enabling high-quality image generation from text prompts~\cite{goodfellow2020generative,karras2019style,rombach2022latent,ramesh2021zero,saharia2022photorealistic,ramesh2022hierarchical}, their extension to 3D remains an ongoing challenge. Diffusion-based methods~\cite{ho2020denoising} have also improved realism in image synthesis, even supporting controls such as image-based guidance and structured constraints like depth, sketches, and key poses~\cite{zhang2023adding, mou2024t2i}. However, these techniques still focus on 2D outputs rather than 3D assets required for AR.

Generating high-quality 3D content is significantly more complex than image synthesis, requiring solutions that balance efficiency and realism. Early text-to-3D models, such as \textit{DreamFusion}~\cite{poole2022dreamfusion}, required over 30 minutes on a powerful GPU~\cite{he2023t3bench} to generate a single asset, making them impractical for in-situ use. \rev{As an alternative, prior systems like LLMR~\cite{llmr-delatorre} relied on large asset libraries (\textit{e.g.,} \textit{Sketchfab}), which---while expansive---often lack imaginative content such as a ``\textit{two-headed giraffe}'' (P2), limiting creativity. Today, techniques aim to accelerate 3D asset creation,} including zero-shot generation~\cite{jain2022zero} and single-image-to-3D approaches~\cite{xu2024instantmesh,liu2024one,boss2024sf3d,hong2023lrm,chen2024lara}. Among these, \textit{InstantMesh}~\cite{xu2024instantmesh} enables rapid 3D lifting (\textit{i.e.,} reconstructing a 3D shape from a 2D image) and texturing from a single image in seconds. To ensure fast and flexible content generation, \sysname{} employs \textit{DALL-E 2}~\cite{ramesh2022hierarchical} to synthesize a 2D image from speech input, then lifts it into 3D using InstantMesh. This pipeline generates a fully textured 3D model in approximately 30 seconds---substantially faster than prior methods in our technical evaluation, and sufficient to support creative iteration in our user study. \rev{Generation speed remains a challenge, but} 3D generative models are rapidly improving in both \rev{speed and fidelity~\cite{Xie2024, xiang2024structured3dlatentsscalable, xiang2025trellis, zhao2025hunyuan3d20}.} As these models advance, our pipeline can adopt faster or higher-quality components---like replacing InstantMesh---without major system changes.

\subsection{AI Assistance in Creative Workflows}
\label{sec:ai-assistance-rel-work}
As a fully functional AI-infused AR authoring tool, \sysname{} presents a unique opportunity for examining how AI can support creative expression in immersive, real-world environments. While we allow free-form use in our study, we also include a controlled investigation of varying levels of AI involvement to examine trade-offs between automation and human agency---a longstanding concern in HCI~\cite{mixedInitiativeUserInterfaces-horvitz,humanAiGuidelines-amershi,hcai-shneiderman}. Prior work shows that while AI can enhance expressivity and efficiency, excessive automation may reduce user control or creative ownership~\cite{ikeaEffect-norton,hcai-shneiderman}. Although this tension has been studied in writing, design, and programming~\cite{nextPhraseSuggestions-bhat,llmsSoftwareTutorialWriters-bhat,metaWriterAIWritingSupport-sun,understandingCreatingArtAi-cetinic,guo2024pen}, its role in AR authoring remains underexplored. Our study helps fill this gap, uncovering not only what users want to create with \sysname{} but also how AI can best assist them along the way.

\section{Design Goals for AI-Infused AR Authoring}
Our research is motivated by an overarching belief that AR authoring tools should allow \textit{anyone} to create \textit{anything}, \textit{anywhere}, removing technical barriers and making immersive content creation as effortless as speaking an idea aloud. Imagine a student in their schoolyard curious about ancient civilizations saying, ``\textit{Construct a Mayan temple next to the swings.}'' and ``\textit{Show a person in historical clothing next to it!}''. After each request, interactive AR content should quickly appear, blending seamlessly into their surroundings. To \rev{pursue} this vision, we synthesized the following design goals:

\textbf{G1: In-Situ AR Authoring Anywhere.} Users should be able to create, modify, and iterate on AR content directly within their environment, treating their surroundings as a canvas for in-situ authoring. \rev{Prior XR authoring systems rely on manually defined and often VR-based environments~\cite{howPeoplePromptVRAuthoring-manesh, dreamCodeVR-giunchi, llmr-delatorre, llmer-chen, enablingWaypointGeneration-fang},} while scene understanding models typically target indoor spaces and require user queries or predefined vocabularies~\cite{takmaz2023openmask3d, peng2023openscene, conceptfusion}. Instead, we need to update these models to \rev{autonomously interpret a wide range of outdoor scenes.}

\textbf{G2: Generate High-Quality 3D Assets Quickly.} To support creativity and maintain flow, users need visually compelling AR assets without long waits. Traditional 3D modeling is time-consuming and technically demanding, and while generative models are improving, they often sacrifice either quality or speed (\textit{e.g.,} \textit{ProlificDreamer}~\cite{wang2023prolificdreamer} takes over 240 minutes on a powerful GPU for a single asset~\cite{he2023t3bench}). Achieving in-situ AR authoring requires generating AR-ready 3D assets in seconds---not minutes or hours.

\textbf{G3: Simple Speech-Driven Interactions.} AR authoring should feel natural and effortless, letting users create and modify scenes with simple voice commands. For example, in the Mayan temple scenario, a student might say, ``\textit{Make the temple bigger}'' or ``\textit{Remove the person.}'' To lower technical barriers, we need LLM-driven speech interactions---enabled by structured scene graphs for spatial context.

\sloppy
\textbf{G4: Adjustable AI Assistance.} AI should support---not override---human creativity, offering just the right level of help while keeping users in control. Preferences for AI involvement vary across users and tasks~\cite{mixedInitiativeUserInterfaces-horvitz,humanAiGuidelines-amershi,hcai-shneiderman,ikeaEffect-norton}. Additionally, when AI makes mistakes, users need clear ways to recover---such as re-prompting or direct manipulation. To support both flexibility and error recovery, AR authoring systems should let users decide how much AI assistance they want and when\rev{, and provide manual tools.}

\section{The \sysname{} System}
\label{sec:system}
\rev{Our goal is to explore how AI can help users bring their ideas to life.} To support this, we developed \textit{\sysname{}}, a novel AI-assisted AR authoring tool that lets users create, \rev{arrange,} and modify virtual content in \rev{diverse, static} real-world environments using speech.

The \sysname{} system consists of three key components: (1) an offline scene processing module, (2) a remote asset generation server, and (3) a mobile AR interface. The scene understanding pipeline structures the environment into a scene graph---a compact textual representation of object labels and their 3D bounding box coordinates. When users request content that is not yet available, the server generates 3D assets on demand. The mobile interface lets users issue speech commands, adjust content manually, and visualize their ideas in-situ. At a high level, \sysname{} retrieves the relevant scene graph, processes voice commands, interprets user intent, fetches or generates 3D assets as needed, updates the scene graph accordingly, and renders changes in the AR scene. We include all LLM and VLM prompts in the Supplementary Materials.

\subsection{Offline Scene Understanding}
To support in-situ AR authoring \rev{\textit{nearly}} anywhere (\textit{Design Goal 1}), we update an open-vocabulary 3D instance segmentation model to operate autonomously in outdoor environments.

\subsubsection{Background Information} \hfill \\
We first introduce \textit{scene graphs} and the \textit{OpenMask3D} model~\cite{takmaz2023openmask3d}, which serve as the foundation of our system.

\textbf{What is a Scene Graph?} A scene graph is a structured textual representation of a visual scene, encoding semantic details such as object labels, locations, and spatial extents. This compact format is well-suited for processing by LLMs. Unlike prior approaches like \textit{ConceptGraphs}~\cite{gu2024conceptgraphs}, which include explicit relationship nodes (\textit{e.g.,} ``\textit{next to}'' or ``\textit{on top of}''), our scene graphs focus solely on individual objects and their spatial properties. Modeling inter-object relationships is left as future work.

\textbf{What is OpenMask3D?} Our scene understanding method builds on OpenMask3D, a state-of-the-art system for open-vocabulary 3D instance segmentation. Given an input point cloud and an RGB-D video with camera poses, OpenMask3D operates in two stages. First, the \textit{Class-Agnostic Mask Proposal (CAMP)} network generates a pool of 3D binary masks, $S_I$, where each mask represents a potential object instance by marking its corresponding 3D points in the point cloud with a value of 1. Second, a \textit{CLIP}~\cite{radford2021learning} embedding is computed for every mask $M \in S_I$. The system performs a depth-based visibility check to identify frames where $M$ is highly visible. Visible points from these frames are used to prompt the \textit{Segment Anything Model (SAM)}~\cite{Kirillov2023segment_anything} at multiple scales, extracting image regions depicting $M$. These regions are then fed into CLIP to generate embeddings, which are aggregated into a single vector per $M$. At test time, users can query objects via text prompts, which are converted into CLIP embeddings and matched against the precomputed embeddings of all masks to retrieve relevant object instances. Notably, OpenMask3D was trained and evaluated primarily on indoor datasets such as \textit{ScanNet}~\cite{dai2017scannet}.

However, we identified two main limitations for our use case. First, the CAMP module often produces an excessive number of masks---frequently over 120---making it difficult to construct compact scene graphs that can be efficiently processed by LLMs. Second, relying on user-defined prompts during use introduces latency, as each query must be embedded and compared against the full set of mask embeddings. Precomputing scene graphs with a predefined vocabulary can avoid this cost but requires a comprehensive label set, which is difficult to define given the variability of outdoor scenes. Hence, OpenMask3D needs to be updated for outdoor AR.

\subsubsection{Our Process} \hfill \\
We now describe our offline scene understanding pipeline, including how we capture point cloud data and adapt OpenMask3D to address key limitations. We also discuss the scalability of our approach.

\begin{figure*}[ht]
    \centering
    \includegraphics[width=\textwidth]{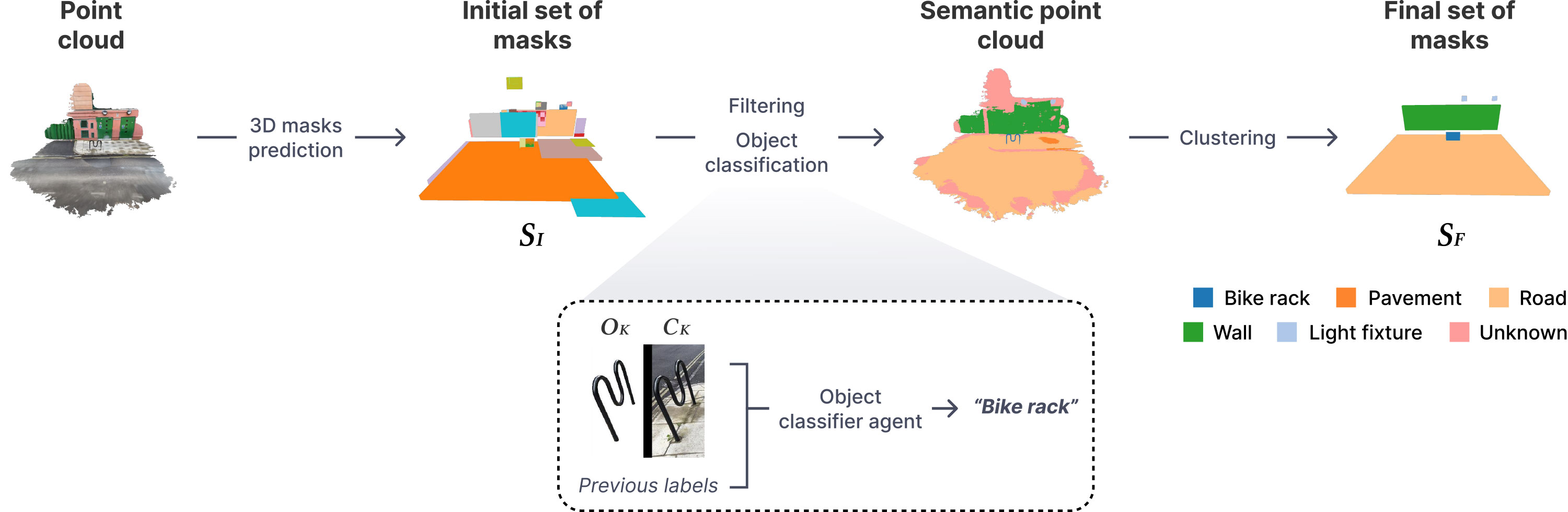}
    \caption{Diagram of the 3D scene understanding pipeline. Given an input point cloud, we first estimate 3D masks. Next, we assign a semantic label to each mask using a VLM and propagate the label to all points within the mask, producing a semantic point cloud. We then cluster nearby points with the same label to infer the final set of 3D masks, from which we extract 3D bounding boxes. For visualization, we show only the bounding boxes, not the underlying masks. The \textit{Pavement} box is enclosed within the \textit{Road} box and is therefore not visible.}
    \label{fig:object_catalog_pipeline}
\end{figure*}

\textbf{Scene capture.}
A key design choice in \sysname{} is to rely on pre-scans of environments and process them offline, rather than running scene understanding models in real-time as users actively scan their surroundings. We chose this approach for three key reasons: first, it enhances ease of use, as live scene understanding requires users to manually and thoroughly scan their environments, introducing unnecessary friction. Instead, digital twins enable precomputed scene understanding, allowing instant retrieval of scene graph data relevant to the user’s location. Second, because users cannot be expected to scan every detail, live scene analysis often results in incomplete context. In contrast, pre-scanned environments can offer a more comprehensive spatial understanding---enabling interactions like placing objects behind the user or real-world structures, even if those areas were never in the camera view. Lastly, real-time scene understanding models typically perform worse than offline methods, especially in complex outdoor environments. \rev{That said, relying on pre-scans may limit scalability compared to live methods and may not reflect dynamic scene changes (\textit{e.g.,} a chopped-down tree or moving people), which we discuss later.}

To generate a 3D representation of a scene, we capture the environment using a commercial depth-sensing device. In our experiments, we used an iPhone 13 Pro, which has LiDAR, running our custom-built scanning app that records RGB images, depth maps, and camera poses. These data sources are integrated into a 3D point cloud, similar to commercial applications like \textit{Scaniverse}~\cite{Scaniverse} and \textit{Polycam}~\cite{Polycam}. Our method is device-agnostic and can be extended to Android devices running \textit{ARCore}~\cite{ARCore}.

\textbf{Pre-Processing.}
To ensure accurate scene understanding and protect user privacy, we apply several pre-processing steps to refine captured data. Personally identifiable information (PII), such as faces and license plates, is removed using an off-the-shelf blurring model~\cite{raina2023egoblur}. We also enhance depth maps by filling holes (\textit{i.e.,} missing values) using a monocular depth model~\cite{yang2024depth_anything_v2}. Because some regions lack depth due to sensor limitations, we infer relative monocular depth and re-scale it with valid LiDAR points to produce dense metric depth maps.

\textbf{Initial Mask Prediction.} 
We use the pre-trained CAMP network from OpenMask3D to generate an initial pool of binary masks, $S_I$, where each mask represents a potential object or object part. However, we observed some masks are small or redundant. Thus, we filter the pool by removing small and duplicate masks, and merging highly overlapping ones, resulting in a refined subset $S_M$.

\textbf{Mask Classification.} 
In this step, we infer a semantic label for each mask in $S_M$. OpenMask3D’s CLIP-based strategy requires either generating scene graphs at test time (via user prompts) or using predefined vocabularies. In contrast, we classify each detected object using a vision-language model (VLM)~\cite{gu2024conceptgraphs}. We modify OpenMask3D’s frame selection strategy to select the image with the highest visibility of the object mask, using monocular depth maps to assess point visibility. From this image, we extract two crops: (1) a context crop ($\mathcal{C}_k$), which includes surrounding scene details, and (2) an object crop ($\mathcal{O}_k$), which isolates the object. These crops are computed only at OpenMask3D’s largest scale to better capture contextual information. We leverage \textit{GPT-4o}~\cite{GPT4o} as the VLM to infer a semantic label from $\mathcal{O}_k$ and $\mathcal{C}_k$, incorporating a running list of previously predicted labels to enforce consistency. This reduces synonym mismatches (\textit{e.g.,} standardizing ``\textit{road}'' instead of allowing similar variations like ``\textit{road surface}''). We refer to this AI agent as the \textit{Object Classifier}, responsible for generating structured semantic labels across diverse outdoor scenes (Figure~\ref{fig:object_catalog_pipeline}).

\textbf{Semantic Point Cloud and Clustering.}
After assigning semantic labels to instance masks, we generate a structured scene representation by storing 3D bounding boxes enclosing each mask in $S_M$. However, $S_M$ may still contain multiple masks for the same object, especially when overlapping masks do not meet the threshold for the prior filter. This redundancy can introduce duplicate instances in the final scene graph.
To address this, we compute a final refined set of masks, $S_F$, using semantic information from a VLM. For each mask $k$ in $S_M$, we propagate its semantic label to its associated 3D points, producing a semantic point cloud. Points not assigned to any mask are labeled as \textit{unknown} and excluded from the final output. We then apply \textit{HDBSCAN}~\cite{mcinnes2017hdbscan} to cluster nearby points with the same label. This merges spatially close, semantically identical masks (\textit{e.g.,} object parts), producing a more compact set $S_F$ compared to $S_M$.
For example, in Figure~\ref{fig:object_catalog_pipeline}, the number of instances is reduced from 208 ($S_I$) to 15 ($S_M$) and finally to 6 ($S_F$).

\begin{figure*}[htbp]
    \centering
    \includegraphics[width=0.98\textwidth]{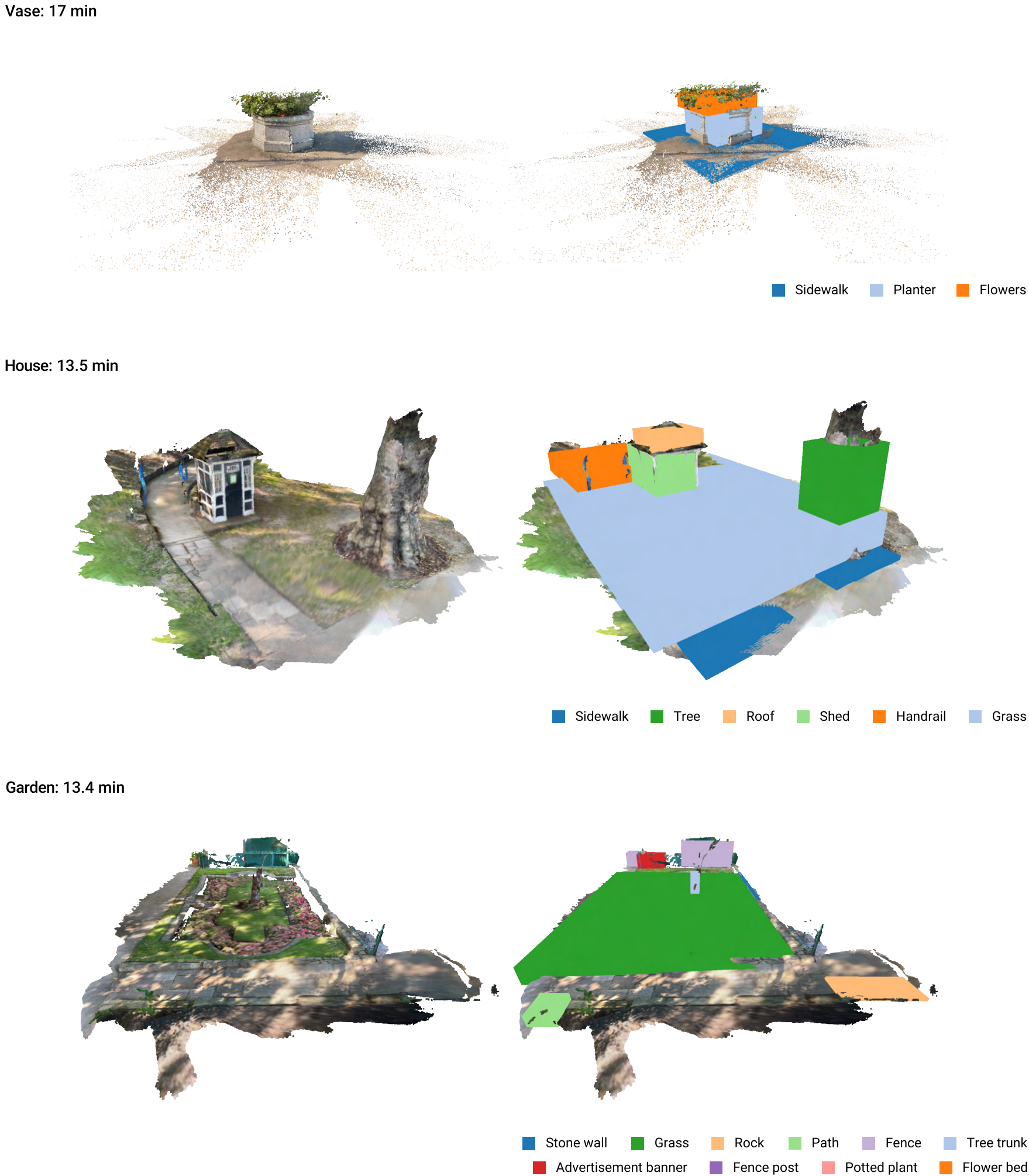}
    \caption{Results of the 3D scene understanding module. For each of the three scans---\textit{Vase}, \textit{House}, and \textit{Garden}---we visualize the input point cloud (left) and the final set of labeled 3D bounding boxes inferred by our scene understanding pipeline (right). We also report the total time (in minutes) required to estimate the scene graph for each scan. Note that some bounding boxes may be enclosed within others and may therefore be occluded.}
    \label{fig:scene_understanding_results}
    \Description[Results of the 3D scene understanding module.]{Results of the 3D scene understanding module.}
\end{figure*}

\textbf{Scene Graph Creation and Deployment.} 
We construct a scene graph by storing semantic labels along with the minimum and maximum values of the 3D axis-aligned bounding boxes enclosing masks in $S_F$. Since these graphs primarily encode static objects, they remain valid across multiple AR sessions and users, as transient elements (\textit{e.g.,} moving people) are typically absent from traditional point cloud reconstructions.
Scene graphs are generated offline using a machine with an NVIDIA L4 GPU; while there is room for optimization, the full pipeline still completes in just a few minutes per scan (Figure~\ref{fig:scene_understanding_results}). During live use, precomputed graphs allow LLM agents to understand the user’s surroundings. Tools like \textit{Niantic’s Visual Positioning System (VPS)}~\cite{VPS} can estimate a user’s precise position relative to the scene graph. For this study, we manually captured all scenes. However, we believe our offline scene understanding pipeline could scale to large pre-scanned datasets already available through platforms like VPS, \textit{Google Street View}~\cite{Streetview}, and \textit{Geospatial API}~\cite{googleGeospatial}. For instance, Niantic VPS currently supports over 1 million scanned locations~\cite{nianticVpsLocations}. Leveraging such resources would enable scalable deployment of \sysname{}.

\subsection{Dynamic Asset Generation}
Running 3D generation models directly on mobile devices is computationally prohibitive. To enable fast AR asset creation (\textit{Design Goal 2}), we deploy a private web server that generates 3D models remotely based on users' speech commands. For example, a user might say, ``\textit{Place a dragon perched on the lamppost},'' prompting the server to return a corresponding textured mesh of a dragon.

\begin{figure*}[htbp]
    \centering
    \includegraphics[width=\textwidth]{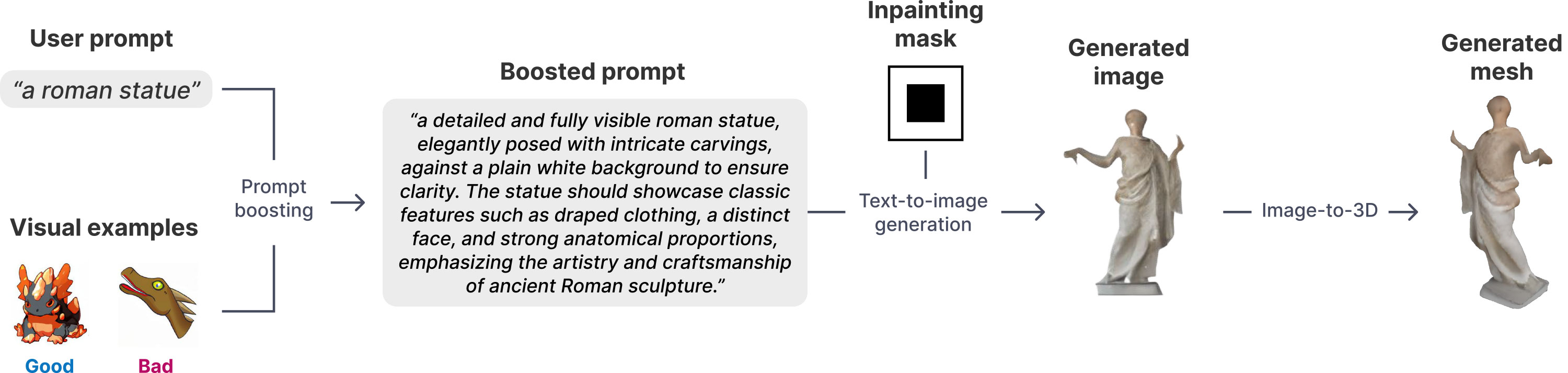}
    \caption{Example of 3D asset generation. Given a user prompt, we first apply prompt boosting, then use Dall-E 2~\cite{ramesh2022hierarchical} to generate a consistent image by editing the center region of a white canvas. The image is then lifted to 3D using InstantMesh~\cite{xu2024instantmesh}. The `Bad'' example (right) illustrates a failure case because it would produce a partial 3D object (\textit{i.e.,} only the dragon’s head). Prompt boosting helps avoid such incomplete generations.}
    \label{fig:asset_generation}
\end{figure*}

To generate assets, we first use a text-to-image model to synthesize an initial image, then apply \textit{DIS}~\cite{qin2022} to segment the foreground subject from the background. While any text-to-image model can be used, image quality does significantly impact the resulting 3D mesh. Images with complex backgrounds, occlusions, or flat perspectives often produce unrealistic models. To address this, we enhance user prompts using \textit{GPT-4o mini}~\cite{GPT4omini}, which expands them with clarifying keywords (\textit{e.g.,} ``\textit{white background}'') to improve visual clarity and depth. We also provide the model with examples of good and bad images. This step---\textit{prompt boosting}---helps \rev{guide} the generated images to meet the requirements for reliable 3D reconstruction. To further improve quality, we instruct \textit{Dall-E 2}~\cite{ramesh2022hierarchical} to edit only the central region rather than generate the full image, encouraging a fully visible, well-defined subject suitable for meshing. We then use \textit{InstantMesh}~\cite{xu2024instantmesh}, an efficient single-image-to-3D model, to lift the image into a fully textured mesh. Because asset generation relies on external services, occasional outages may occur. In such cases, we fall back to the original user prompt (without boosting) or switch to \textit{Stable Diffusion Turbo}~\cite{sauer2023adversarial} as a local text-to-image generator. Figure~\ref{fig:asset_generation} illustrates the full pipeline.

\subsection{Real-World User Authoring}
\label{sec:mobile-app}
To support seamless in-situ AR authoring (\textit{Design Goal 3}), we developed a mobile interface that enables speech-driven interactions with advanced AI models. We built it using \textit{Unity} 2022.3.33f1\footnote{\url{https://unity.com}}, \textit{ARFoundation} 5.1.4~\cite{ARFoundation}, and \textit{Niantic Lightship ARDK} 3.5.0~\cite{ARDK}.

\begin{figure*}[htbp]
    \centering
    \includegraphics[width=\textwidth]{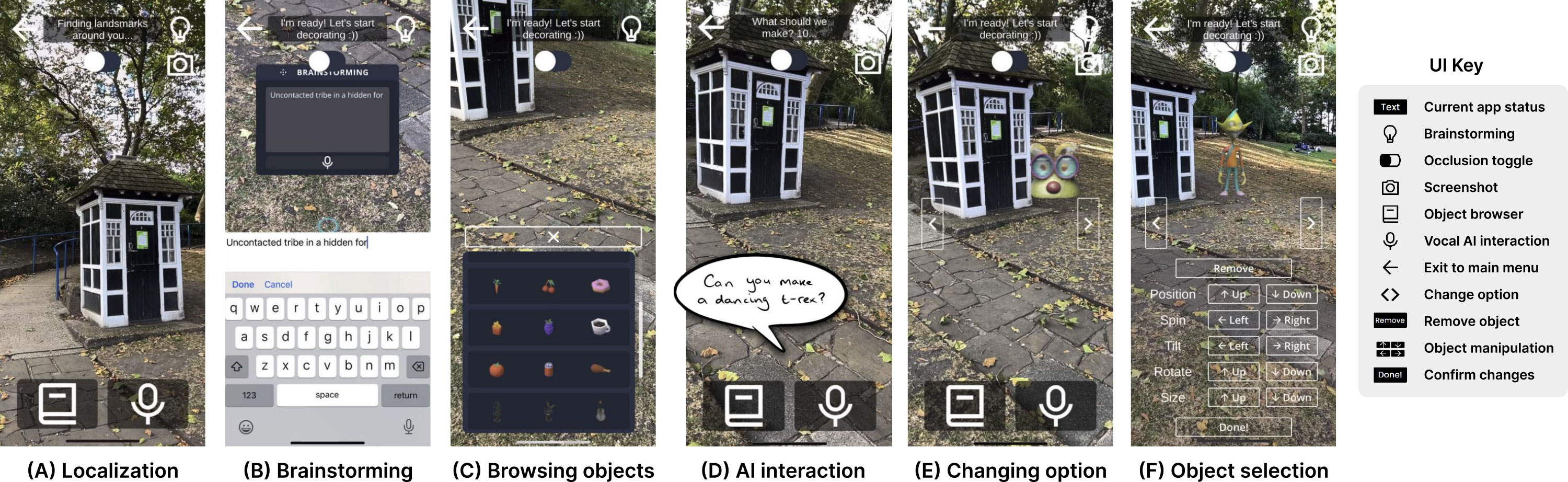}
    \caption{Different screen captures of the \sysname{}'s mobile interface showing the UI layout and functionalities. Users can access manual, AI-assisted, and AI-decided modes across different features through buttons on the screen.}
    \label{fig:mobile_app_overview}
    \Description[The UI elements in the \sysname{} app.]{The UI elements in the \sysname{} app.}
\end{figure*}

We designed \sysname{} to support five core interactions for authoring an AR scene: \textit{brainstorming}, \textit{model creation}, \textit{placement}, \textit{editing}, and \textit{removal}. For each task, users can choose from three levels of AI involvement (\textit{Design Goal 4}): ``\textit{manual}'', where they maintain full control; ``\textit{AI-assisted}'', where the system offers multiple suggestions; and ``\textit{AI-decided}'', where AI autonomously executes the task and presents a single best option. To facilitate these interactions, \sysname{} employs three specialized LLM agents: a \textit{Brainstorming} agent for idea generation, an \textit{Action Plan} agent for interpreting user requests and structuring tasks, and an \textit{Assembly} agent for executing actions \rev{like asset placement. AI-assisted and AI-decided modes share this LLM pipeline but differ in autonomy and how results are presented to the user.} Figure~\ref{fig:mobile_app_overview} illustrates the interface and supported interactions.

\textbf{Localization.} 
Users begin by pointing their phone around to localize to a nearby \textit{Point of Interest (POI)}---a geotagged location---using Niantic’s VPS. Once the system determines the user’s position, it retrieves the corresponding precomputed scene graph, providing a structured representation of its surroundings for the LLM agents. \sysname{} then displays: ``\textit{I'm ready! Let's start decorating!}''

Our system updates the retrieved scene graph to reflect the evolving AR experience. As users request new virtual content, it is added to the local scene graph. Each object has a unique identifier (GUID), a name, and position, rotation, scale, and bounding box dimensions in Unity’s world coordinate system. The graph also includes an on-screen visibility flag for handling spatially ambiguous queries (\textit{e.g.,} ``\textit{Place the T-Rex here}'') and an action tag to track LLM-assigned modifications awaiting execution. Together, this structure provides essential context for iterative, LLM-driven interactions.

\textbf{Brainstorming Ideas.} 
Before editing the AR scene, users can brainstorm using a post-it-style interface triggered by the light bulb button. They can type ideas manually or ask AI for suggestions---either in a single prompt (AI-decided) or through back-and-forth conversation (AI-assisted). When speaking to the Brainstorming agent, \sysname{} captures audio using Unity’s microphone\footnote{\url{https://docs.unity3d.com/ScriptReference/Microphone.html}}, transcribes it with \textit{Whisper}~\cite{Whisper}, and prompts GPT-4o along with the current scene graph to ground ideas in the user’s AR environment. The post-it window is movable to prevent visual obstruction and can be closed by tapping the button again.

\textbf{Creating 3D Assets.} 
Users can add virtual content by selecting from a preset library or asking the AI to generate new assets. For manual selection, tapping the book button in the bottom left opens a scrollable grid of virtual objects. For AI-driven creation, users tap the microphone button and describe what they want. The system returns the top result (AI-decided), with optional left and right arrows to browse alternatives (AI-assisted). To support AI-assisted creation, \sysname{} runs three asset generators in parallel, each producing a distinct asset aligned with the user’s request.

If AI creation is used, \sysname{} transcribes the user’s speech and sends it---along with the current scene graph---to the \textit{Action Plan} agent. This agent assigns each virtual object an action tag: (1) \textit{none} (no change), (2) \textit{remove}, (delete from the scene), (3) \textit{update} (modify properties like position, rotation, or scale), (4) \textit{create\_resources} (instantiate a preset model), (5) \textit{create\_persistent} (load a previously generated model), or (6) \textit{create\_new} (request a new mesh from the remote asset generation server). \sysname{} then either retrieves an existing model (\textit{create\_resources}, \textit{create\_persistent}) or generates a new one remotely (\textit{create\_new}). The assets are added to the scene to compute spatial properties like bounding box dimensions.

\textbf{Arranging Virtual Content.} 
Users can place, modify, and remove virtual objects either manually or with AI tools. For manual placement, users tap the `Place Object' button to position a selected model at the blue visual indicator, which marks where a ray from the center of the screen intersects ARDK’s live mesh~\cite{Meshing} (\textit{i.e.,} the estimated geometry of the real world). Tapping on a placed object opens an editing window for adjusting position, rotation, and scale (manual modification) or deleting the object (manual removal).

In AI mode, users can issue verbal commands such as ``\textit{Put a silly hat on the statue.}'' The \textit{Assembly} agent interprets action tags assigned by the Action Plan agent and determines how to arrange content. \rev{The Assembly agent uses each object's transform, along with its minimum and maximum bounds (computed via a \textit{BoxCollider}), for spatial reasoning---such as aligning the top of a statue with the base of a silly hat or scaling a T-Rex to appear larger than nearby objects. It determines each object's placement, rotation, and scale to make it look realistically situated in the real-world scene. The agent then} performs AI-decided placement (for \texttt{create\_resources}, \texttt{create\_persistent}, and \texttt{create\_new} tags), modification (\texttt{update}), or removal (\texttt{remove}), displaying the top result by default. Users can use the left and right arrows to browse alternative placement, modification, or removal options (AI-assisted mode), generated by three parallel Assembly agent (LLM) calls.

\textbf{Example AI Creations.} During both the technical evaluation and user study, users had access to the full set of features. Figure~\ref{fig:proof_by_demonstration} showcases AR scenes authored by the research team, while Figure~\ref{fig:user_creations} highlights participant-created scenes, often composed using a mix of AI and manual tools. To isolate the performance of \sysname{}’s AI components, we also captured examples generated entirely by AI---without any manual input from participants---in Figure~\ref{fig:ai-generations}.

\section{Technical Evaluation}
\sloppy

\begin{figure*}[!ht]
    \centering
    \includegraphics[width=\textwidth]{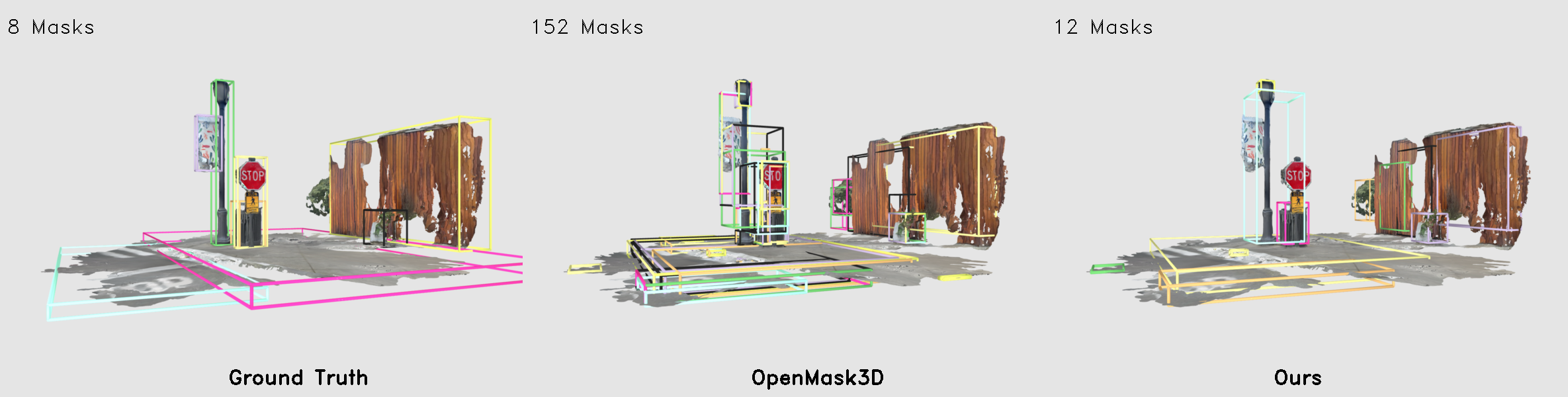}
    \caption{From left to right: bounding boxes from the ground truth, OpenMask3D~\cite{takmaz2023openmask3d}, and our proposed method. OpenMask3D predicts a large number of masks, resulting in excessive bounding boxes that over-represent the same scene objects. In contrast, our method produces fewer, more accurate boxes. (Box colors are arbitrary and can be ignored.)}
    \label{fig:benchmark-comparison}
\end{figure*}

We conducted a technical evaluation of \sysname{} to assess the performance of its core components. First, we measured component-level latency to evaluate its feasibility for in-situ, real-time authoring (Table~\ref{tab:timings}). Across 50 trials, our system averaged $33.92 \pm 5.83$ seconds---substantially faster than prior systems like LLMR~\cite{llmr-delatorre}, which reports $90.98 \pm 24.88$ seconds in an empty VR scene and $49.16 \pm 7.87$ seconds in a virtual bathroom, \rev{though with the caveat that its latency primarily stems from iterative refinement, whereas ours is due to asset generation.} Next, we compared our two key technical contributions---scene understanding and asset generation---against state-of-the-art baselines. Finally, we conducted a proof-by-demonstration to illustrate that \sysname{} can scale across diverse outdoor environments.

\begin{table}[h!]
\caption{Latency analysis of key components in \sysname{}. We report mean $\pm$ standard deviation (in seconds) for each pipeline step, averaged over 50 trials.}
\footnotesize
\begin{tabular}{ l l }
\toprule
 \textbf{Component} & \textbf{Time} \\ 
 \midrule
 Prompt Boosting & 2.53s $\pm$ 0.91s \\ 
 \midrule
 Image Generation & 12.53s $\pm$ 2.48s \\ 
 \midrule
 Background Removal & 0.04s $\pm$ 0.002s \\ 
 \midrule
 Image to Mesh & 9.14s $\pm$ 0.08s \\ 
 \midrule
 In-App LLM Agents & 9.68s $\pm$ 1.24s \\ 
 \midrule
 Total & 33.92 $\pm$ 5.83s\\ 
\bottomrule
\end{tabular}
\label{tab:timings}
\end{table}

\subsection{Scene Understanding Pipeline}
We evaluated our scene understanding pipeline on five distinct outdoor scenes. Because existing outdoor benchmarks primarily focus on driving scenarios~\cite{Geiger2012CVPR,caesar2020nuscenes}, they are unsuitable for our purposes. We therefore captured our own data and generated ground truth scene graphs by manually labeling each scene. Each node in a graph represents an object as a 3D bounding box and a human-defined semantic label. One member of the research team performed the initial labeling, and two others reviewed it for bias and accuracy.

To create these ground truth graphs, we developed a custom annotation tool that loads point clouds and allows users to brush over points using different colors and brush sizes. This lets users assign a unique color to each object and define its semantic label, producing a structured scene graph. Using this dataset, we evaluated how well different methods detect and describe objects. To compute metrics, we used the Hungarian algorithm to match predicted bounding boxes to ground truth boxes based on Intersection over Union (IoU). A match was counted as a true positive if IoU $\geq 0.25$.

We report the following metrics: mean Recall, computed as the average per-scene Recall (true positives over ground truth instances), and mean Semantic Similarity (mean SS), the average cosine similarity between CLIP~\cite{radford2021learning} embeddings of ground truth and predicted labels for true positives. We also report total predicted masks (N) per method. Across all five scenes, there are 27 ground truth instances. Experiments using GPT-4o were repeated five times with $\mathrm{top_p}=0.1$ using the latest available model.

\begin{table*}[hbt!]
\footnotesize
    \centering %
    \caption{Evaluation of 3D scene understanding pipelines. We build on OpenMask3D~\cite{takmaz2023openmask3d} to produce more compact scene graphs. The benchmark includes five manually labeled scenes with 27 total ground truth bounding boxes. We report the total number of predicted masks (N), mean Recall, and mean Semantic Similarity (mean SS) using a 0.25 IoU threshold. Rows A–C are ablations: (A) adds GPT-4o labeling and initial mask filtering, (B) incorporates monocular depth, and (C) uses CLIP~\cite{radford2021learning} instead of GPT-4o. GPT-4o results are averaged over five runs and reported as mean ± standard deviation.}

    \begin{tabular}{l cccc llll}
        \toprule
        & \multicolumn{4}{c}{\textbf{Components Used}} & \multicolumn{1}{c}{} & \multicolumn{2}{c}{\textbf{Evaluation Metrics}} \\
        \cmidrule(lr){2-5} \cmidrule(lr){7-8}

        Method & Filtering & Monocular Depth & Labeling & Clustering & \shortstack{N} &  \shortstack{mean Recall} $\uparrow$ & \shortstack{mean SS} $\uparrow$ \\
        \midrule

        OpenMask3D \cite{takmaz2023openmask3d}   &  & & CLIP &  & 752 & 0.800 & 0.738 \\
        \hline
        Ablation A    &      \checkmark      &  & GPT-4o &  &  59 & 0.508 & 0.659 ($\pm$ 0.008) \\
        Ablation B   & \checkmark & \checkmark &  GPT-4o & & 60 & 0.558 & 0.791 ($\pm$ 0.010) \\
        Ablation C   & \checkmark & \checkmark & CLIP & & 60   & 0.558 & 0.730 \\
        \hline
        \textbf{Ours} &   \checkmark  &    \checkmark  & GPT-4o &  \checkmark  & 49 ($\pm$ 1) & 0.622 ($\pm$ 0.087) & 0.791 ($\pm$ 0.073) \\

        \bottomrule
    \end{tabular}
    \label{tab:scene-understanding}
\end{table*}

Table~\ref{tab:scene-understanding} ablates variants of the scene understanding pipeline. The first row reports OpenMask3D~\cite{takmaz2023openmask3d} results using a 4,500-class vocabulary from~\cite{zhang2023recognize} to assign a label to each detected mask. OpenMask3D shows strong recall, but the large number of predicted masks suggests many may be redundant, creating distractors for LLM agents. Ablation A replaces CLIP with GPT-4o and adds a filtering step to reduce the number of masks. While this lowers the total number of masks, it also reduces the number of correctly predicted masks and semantic label quality. Ablation B incorporates dense monocular depth in metric scale, improving both recall and semantic similarity---suggesting that better visibility yields more accurate crops. Ablation C reintroduces CLIP on the same inputs as B but produces lower semantic scores, indicating that GPT-4o yields more accurate labels. Finally, our full method adds a clustering step to merge nearby masks with the same label, further reducing redundancy and producing compact yet meaningful scene graphs.

\subsection{Asset Creation with AI}
To evaluate the efficiency and quality of our text-to-3D generation pipeline, we leveraged T$^3$Bench~\cite{he2023t3bench}, a benchmark designed to assess text-to-3D methods across varying scene complexities. T$^3$Bench provides standardized text prompts and computes a quality score based on multi-view 2D renderings generated from 3D input assets. It also includes benchmarking results for state-of-the-art text-to-3D models, including \textit{ProlificDreamer}~\cite{wang2023prolificdreamer}, \textit{MVDream}~\cite{shi2023MVDream}, \textit{DreamFusion}~\cite{poole2022dreamfusion}, and \textit{DreamGaussian}~\cite{tang2023dreamgaussian}.

We report official scores and timings for these methods in Table~\ref{tab:asset-generation} and compare them against our strategy using the single objects generation benchmark. Our method achieves sub-minute generation times---crucial for in-situ AR authoring---while maintaining reasonable visual quality. Although our assets are slightly lower in quality than those from ProlificDreamer and MVDream, \rev{they outperform DreamFusion and DreamGaussian.} However, higher-quality models come at a significant cost: ProlificDreamer requires 240 minutes and MVDream 30 minutes per asset on a powerful GPU, making them unsuitable for real-time AR. In contrast, our approach balances speed and quality, enabling fast asset generation while preserving usability---making it the most practical solution for in-situ AR authoring. As 3D generative models continue to improve in \rev{both speed and fidelity~\cite{Xie2024, xiang2024structured3dlatentsscalable, xiang2025trellis, zhao2025hunyuan3d20},} future work should explore these evolving alternatives.

\begin{table}[h!]
\caption{Benchmark results comparing state-of-the-art text-to-3D pipelines with our approach, evaluated on the T$^3$Bench dataset~\cite{he2023t3bench}. Prior methods are impractical for in-situ AR authoring due to long runtimes. Our approach, combining InstantMesh with Dall-E 2 and prompt boosting, achieves sub-minute generation while maintaining quality.}
\footnotesize
\begin{tabular}{ l| l l } 
\toprule
 \textbf{Model Name} & \textbf{Time} & \textbf{Quality $\uparrow$}\\ 
 \midrule
 DreamFusion~\cite{poole2022dreamfusion} & 30 min & 24.9 \\ 
 ProlificDreamer~\cite{wang2023prolificdreamer} & 240 min & 51.1 \\ 
 MVDream~\cite{shi2023MVDream} & 30 min & 53.2 \\ 
 DreamGaussian~\cite{tang2023dreamgaussian} & 7 min & 19.9 \\
 \leftcell{InstantMesh~\cite{xu2024instantmesh} + Dall-E 2~\cite{dallEEditing-openAi}} & < 1 min & \rev{32.6}\\ 
 \midrule
 \leftcell{InstantMesh + Dall-E 2\\+ Prompt Boosting (Ours)} & < 1 min  & \rev{34.8}\\ 
\bottomrule
\end{tabular}
\label{tab:asset-generation}
\end{table}

\begin{figure*}[ht]
    \centering
    \includegraphics[width=\textwidth]{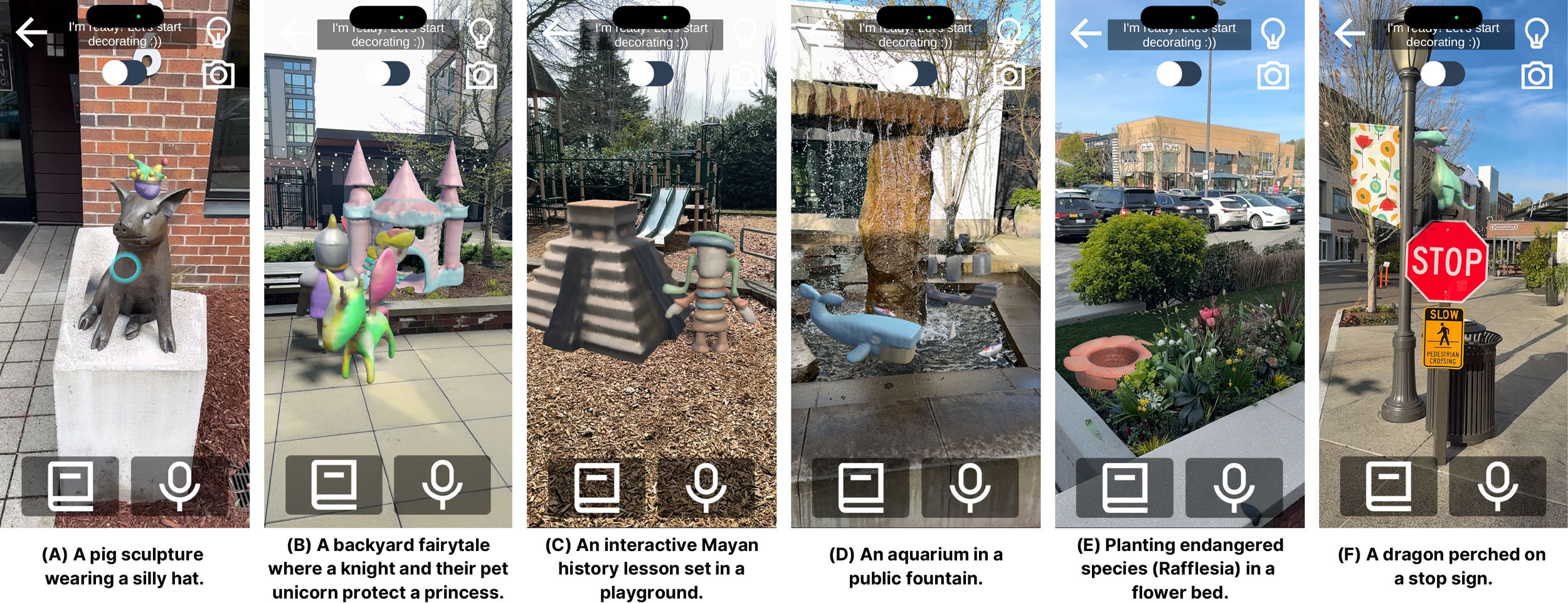}
    \caption{Six example creations from our technical evaluation, situated in a park, schoolyard, playground, shopping center, and backyard. Each scene was first generated with AI tools, then refined with light manual adjustments to reflect typical \sysname{} use. Some are whimsical (A, F), while others are educational (C, E) or playful (B, D).}
    \label{fig:proof_by_demonstration}
\end{figure*}

\subsection{Proof by Demonstration}
To evaluate whether \sysname{} scales across diverse outdoor settings, we conducted a proof-by-demonstration study at 10 Points of Interest (POIs) spanning five distinct sites in two cities. These included statues, flower beds, trees, fountains, play structures, and more. Figure~\ref{fig:proof_by_demonstration} showcases example AR scenes created by the research team using \sysname{}. For instance, we authored a fairytale in a backyard, a Mayan history lesson on a playground, and an aquarium inside a public fountain---demonstrating \sysname{}’s adaptability across varied environments.

\section{User Study}
To complement our technical evaluation, we conducted a three-part within-subjects user study with 4 pilot participants and 20 study participants. This in-situ study took place in a public park and aimed to: (1) explore the types of AR experiences users want to author outdoors, (2) observe how and when users engage with manual and AI-driven features, and (3) identify current limitations and future opportunities in AI-infused AR authoring.

\subsection{Participants}
We recruited participants via mailing lists and snowball sampling, screening them through a demographic questionnaire on age, gender, and experience with 2D/3D creativity tools, AR technologies, and AI chat systems. To be eligible, participants had to be at least 18 years old with no visual or auditory impairments. From 147 respondents, we invited 34 to balance demographic diversity and prior experience; 24 participated in the study (4 in pilot sessions).

Participants ranged from 18 to 61 years old ($M=$35, $SD=$11.8) and identified as 33.3\% female, 58.3\% male, and 8.3\% non-binary. Half had no prior experience with 3D creativity tools, while 25.0\% were slightly familiar, 12.5\% very familiar, and the remainder evenly split between moderately familiar and familiar. In AR, 4.2\% were unfamiliar, 33.3\% moderately familiar, and the rest evenly distributed across slightly familiar, familiar, and very familiar. AI chat systems were more widely used: 12.5\% were familiar, 37.5\% very familiar, and the remainder evenly divided between slightly and moderately familiar. Participants received a £50 gift card for their time.

\begin{table*}[hbt!]
\footnotesize
\caption{The three AI modes and five features (15 total trials) participants engaged with in Part 1 of the study.}
\centering
\begin{tabular}{l l l l}
\toprule
\textbf{Feature} & \textbf{A: Manual} & \textbf{B: AI-Assisted} & \textbf{C: AI-Decided} \\
\midrule
Brainstorming Ideas & \leftcell{User either thinks aloud or writes down ideas\\in the app.} & \leftcell{User converses with LLM to collaboratively\\come up with idea(s). User chooses final idea.} & \leftcell{Single-turn communication with LLM for\\ideation.} \\
\midrule
3D Asset Creation & \leftcell{User searches for and selects 3D assets from\\pre-existing database.} & \leftcell{AI generates three different 3D assets, of\\which the user selects one.} & \leftcell{AI generates and selects a single 3D asset.} \\
\midrule
Object Placement & \leftcell{User moves the cursor by aiming the camera,\\then taps to place the object.} & \leftcell{AI determines three different positions to\\place the object, of which the user selects one.} & \leftcell{AI determines where to position the\\newly-created object.} \\
\midrule
Object Modification & \leftcell{User taps to select the object, then moves \\around and taps buttons to edit its pose.} & \leftcell{User asks LLM to edit the object's pose. AI\\determines three edit arrangements, of\\which the user selects one.} & \leftcell{User asks LLM to edit the object's pose.\\AI chooses the final arrangement.} \\
\midrule
Object Removal & \leftcell{User taps on an object and then taps a button\\to remove it.} & \leftcell{User asks AI to remove object(s). AI shows\\three possibilities, of which the user selects one.} & \leftcell{User asks AI to remove object(s). AI chooses the\\final removal(s).} \\
\bottomrule
\end{tabular}
\label{tab:part1}
\end{table*}

\subsection{Procedure}
Our in-person study took place in a busy public park featuring varied terrain, including grass, pavement, stairs, a shed, and trees. This complex setting allowed participants to interact with diverse real-world objects while testing \sysname{}'s adaptability. Study sessions were recorded, capturing participants’ phone screens and audio for later analysis. We collected both quantitative and qualitative data through surveys and semi-structured interviews, with full study materials available in the Supplementary Materials. Each 2-hour session included an initial tutorial and three study phases:

\textbf{Tutorial.} The session began with participants watching a 5-minute introductory video explaining the study and system features. They then had the opportunity to ask questions before proceeding.

\textbf{Part 1: Comparison Task.} As a novel outdoor AR authoring tool, \sysname{} raises open questions about AI’s role in the authoring process. To explore when and how much AI involvement users preferred, we first conducted a structured comparison before allowing free-form creation. Participants began with a 3-minute overview video before interacting with three system modes: (A) \textit{manual}, where users tapped the screen and physically moved to manipulate the AR scene; (B) \textit{AI-assisted}, where the AI suggested options but users made final decisions; and (C) \textit{AI-decided}, where the AI autonomously generated a single output. They performed five core AR authoring tasks---(1) \textit{brainstorming}, (2) \textit{object creation}, (3) \textit{placement}, (4) \textit{modification}, and (5) \textit{removal}---across all three AI modes, completing 15 trials (1A–5C; see Table~\ref{tab:part1}). Mode order was counterbalanced using a Latin Square. After each trial, participants completed a post-task questionnaire with UMUX-LITE~\cite{Lewis2013}, a two-item usability measure adapted from SUS~\cite{Brooke1996}, and NASA-TLX~\cite{Hart2006} ratings for mental demand, performance, effort, and frustration. At the end of this phase, we asked which mode participants preferred overall and which they would use for additional features such as music, sound effects, animations, event triggers, and object pinning.

\textbf{Part 2: Free-Form Authoring Task.} Beyond structured comparisons, observing how and what users create without researcher intervention is critical---and only possible with a fully functional prototype. In this phase, participants used the full \sysname{} system to freely author AR scenes of their own imagination for 10–30 minutes. Afterward, they completed the Creativity Support Index (CSI)~\cite{Cherry2014} questionnaire and provided qualitative feedback on \sysname{}’s perceived usability and creativity support.

\textbf{Part 3: Brainstorming and Co-Design.} Lastly, we conducted a semi-structured interview to gather insights on participant experiences, preferred features, and ideas for system improvement. We prepared 11 qualitative questions covering what they created, their workflow choices, trade-offs between manual and AI-driven authoring, and desired future enhancements. Follow-up questions were asked based on responses, aiming to identify \sysname{}’s limitations and opportunities for future development.

\subsection{Analysis}
We analyzed data from three sources: questionnaire responses, session observations, and interview transcripts. Quantitative data were examined using a Friedman test, followed by Wilcoxon signed-rank tests with Holm’s sequential Bonferroni correction for pairwise comparisons. Qualitative data were analyzed using reflexive thematic analysis~\cite{Braun2006, Braun2019}. The first author developed an initial codebook, which was refined collaboratively with another researcher. The final codebook comprised 56 codes, applied to 412 participant quotes and reviewed by an additional researcher.

\section{Results}
We first present findings from structured comparison tasks---including perceived usability, task load, and creativity support---to understand how different levels of AI involvement affect AR authoring. Next, we analyze free-form authoring behaviors to offer deeper insight into how users naturally engage with \sysname{} and the types of AR experiences they create. Finally, we synthesize key themes from qualitative feedback, highlighting user preferences, expectations around AI collaboration, and opportunities for designing future AI-powered AR authoring tools. Participant quotes have been lightly edited for clarity and concision.

\subsection{Comparing Levels of AI Involvement}
\begin{figure*}[htbp]
    \centering
    \includegraphics[width=\textwidth]{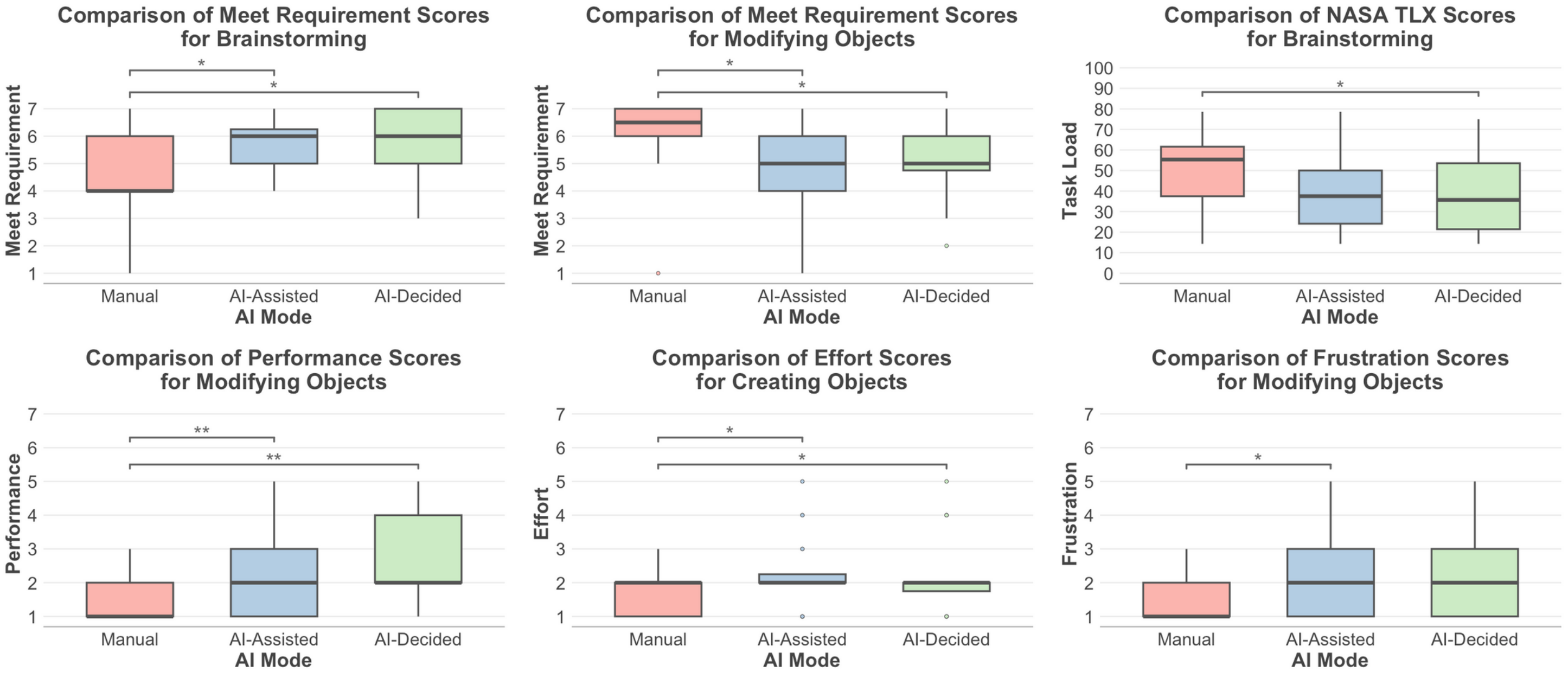}
    \caption{Boxplots of significant results from Part 1 quantitative data. Higher values indicate better outcomes for Meet Requirements, while lower values are better for NASA-TLX, Performance, Effort, and Frustration scores.}
    \label{fig:part1_signif_quant}
\end{figure*}

In Part 1, we quantitatively compared (A) manual, (B) AI-assisted, and (C) AI-decided modes across five core AR authoring tasks: brainstorming, object creation, placement, modification, and removal. Post-trial questionnaires measured usability (UMUX-LITE) and task load (NASA-TLX), with Table~\ref{tab:part1_overall_quant} showing overall results and Figure~\ref{fig:part1_signif_quant} highlighting significant differences. This phase aimed to establish an initial comparison of AI involvement across tasks.

\begin{table*}[htb]
\caption{Usability (UMUX-LITE, on the left) and task load (NASA-TLX, on the right) data collected in Part 1. We report average $\pm$ standard deviation. For statistically significant data, we also provide a plot.}
\footnotesize
\centering
\begin{minipage}{.48\textwidth}
\centering
\small
\begin{tabular}{l c c c}
\toprule
\textbf{Feature} & \textbf{A} & \textbf{B} & \textbf{C} \\
\midrule
Brainstorming Ideas & 66.2 ± 16.4 & 76.8 ± 7.4 & 75.4 ± 10.6 \\
Creating Objects & 69.8 ± 11.2 & 70.8 ± 12.2 & 67.9 ± 13.2 \\
Modifying Objects & 73.0 ± 16.0 & 66.2 ± 14.9 & 68.1 ± 13.5 \\
Placing Objects & 72.2 ± 15.3 & 70.6 ± 10.9 & 67.6 ± 15.9 \\
Removing Objects & 79.2 ± 16.7 & 80.9 ± 8.6 & 79.5 ± 12.4 \\
\bottomrule
\end{tabular}
\subcaption*{UMUX-LITE}
\end{minipage}%
\hspace{0.02\textwidth}
\begin{minipage}{.48\textwidth}
\centering
\small
\begin{tabular}{l c c c}
\toprule
\textbf{Feature} & \textbf{A} & \textbf{B} & \textbf{C} \\
\midrule
Brainstorming Ideas & 48.6 ± 20.3 & 40.2 ± 20.2 & 37.3 ± 18.0 \\
Creating Objects & 27.5 ± 12.3 & 33.8 ± 12.0 & 35.0 ± 12.2 \\
Modifying Objects & 30.5 ± 15.6 & 32.0 ± 15.0 & 34.1 ± 14.9 \\
Placing Objects & 34.3 ± 18.2 & 28.8 ± 12.7 & 33.2 ± 13.2 \\
Removing Objects & 24.1 ± 14.5 & 24.8 ± 13.5 & 23.2 ± 14.5 \\
\bottomrule
\end{tabular}
\subcaption*{NASA-TLX}
\end{minipage}
\label{tab:part1_overall_quant}
\end{table*}

\textbf{Usability.} UMUX-LITE scores showed no significant differences in overall usability across AI modes. However, analyzing individual questions revealed task-specific differences in how well each mode met participants' needs. Friedman tests found significant differences for brainstorming ($\chi^2(2,N=20)=6.58$, $p<0.05$) and object modification ($\chi^2(2,N=20)=13.07$, $p<0.01$), but not for other tasks. Post-hoc Wilcoxon signed-rank tests revealed that AI-assisted ($V=151$, $p<0.05$) and AI-decided ($V=165$, $p<0.05$) modes better met user requirements for brainstorming than manual. Conversely, manual outperformed AI-assisted ($V=11$, $p<0.05$) and AI-decided ($V=23.5$, $p<0.05$) for object modification. For the ease-of-use question, no significant differences were observed across modes.

\textbf{Task Load.} NASA-TLX scores showed a significant difference for brainstorming ($\chi^2(2, N=20)=7.21$, $p<0.05$), with manual mode inducing significantly higher overall task load than AI-decided ($V=21$, $p<0.05$). We also examined the mental demand, performance, effort, and frustration components separately, as these dimensions were particularly relevant to our study.

\textbf{Mental Demand.} No significant differences in mental demand were found across modes, indicating no evidence that any particular mode was more mentally demanding than others.

\textbf{Performance.} Object modification performance differed significantly across modes ($\chi^2(2, N=20)=11.29$, $p<0.01$), with manual outperforming AI-assisted ($V=82$, $p<0.01$) and AI-decided ($V=88$, $p<0.01$).

\textbf{Effort.} Object creation effort differed significantly across modes ($\chi^2(2, N=20) = 9.14$, $p<0.01$), with manual requiring significantly less effort than both AI-assisted ($V=41.5$, $p<0.05$) and AI-decided ($V=36$, $p<0.05$). When asked why, participants noted that while AI features demanded less active input and decision-making, they still had to wait for system responses---suggesting they equated effort with overall task duration.

\textbf{Frustration.} Frustration during object modification varied significantly ($\chi^2(2, N=20)=8.39$, $p<0.05$), with manual mode causing less frustration than AI-assisted ($V=45$, $p<0.05$).

\textbf{Overall Preference.} After completing all trials, 12 participants preferred manual mode, 10 favored AI-assisted, and 2 equally preferred both (P5, P16). Participants appreciated the manual mode for its control (10/20) and precision (9/20), helping them create scenes that more precisely matched their vision. AI-assisted was valued for fostering creativity (6/20) and offering multiple AI-generated options for review (5/20). AI-decided was least favored, as participants found it ``\textit{too rigid and deterministic}'' (P5), though some acknowledged its ability to quickly generate results (4/20) and reduce the mental effort of decision-making (4/20).

\begin{figure}[h!]
    \centering
    \includegraphics[width=.85\linewidth]{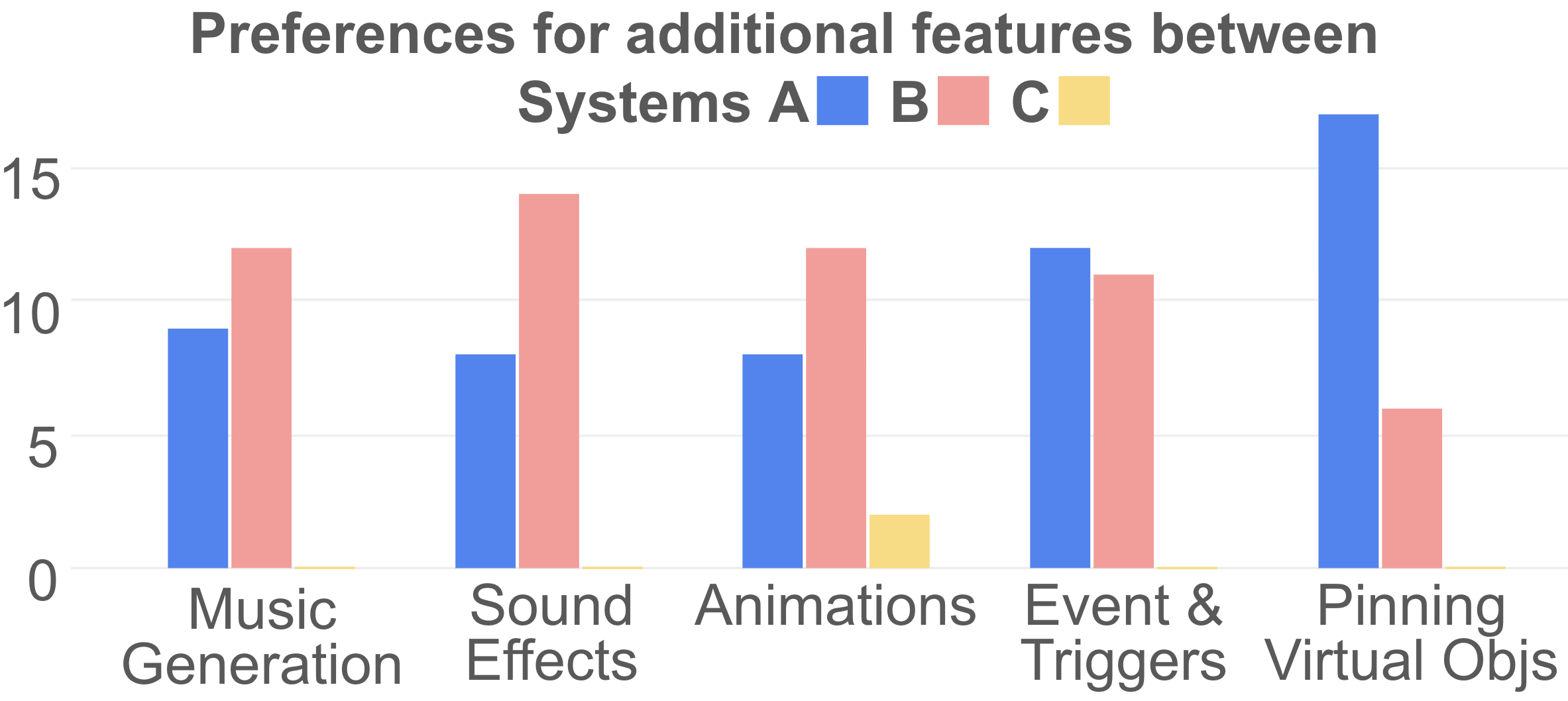}
    \caption{A bar graph showing participant preferences for level of AI involvement across proposed additional features.}
    \vspace{-10px}
    \label{fig:additional_feature_preferences}
\end{figure}

\textbf{Authoring Preferences for Additional Features.}
Participants proposed future features and indicated their preferred AI mode for each, including background music, sound effects, animations, event triggers, and object pinning. Preferences are summarized in Figure~\ref{fig:additional_feature_preferences}. Overall, participants favored manual mode for tasks requiring fine-grained control, such as pinning objects to specific parts of real-world surfaces, and preferred AI-assisted mode for creative, generative tasks like adding sounds and animations.

\textbf{Summary.} While AI-assisted mode was expected to be the most preferred, participants’ preferences varied across tasks due to trade-offs between speed, creativity, and precision. AI-assisted was appreciated for generating creative options with less decision-making, but manual mode was valued for precise adjustments, such as fine-tuning object placement, despite requiring more active input and time. AI-decided was helpful for brainstorming but lacked the control needed for tasks driven by specific user intent. These findings suggest that future AI-powered AR authoring tools should support all three modes, enabling users to adjust automation and control based on their needs at different stages of the authoring process.

\subsection{Free-Form Authoring with \sysname{}}
In Part 2, participants freely authored AR scenes using the full \sysname{} system for 10–30 minutes before providing quantitative and qualitative feedback. Below, we present findings on creativity support, followed by an analysis of what participants created and how they used the system without researcher intervention.

\begin{figure}[h!]
    \centering
    \includegraphics[width=\linewidth]{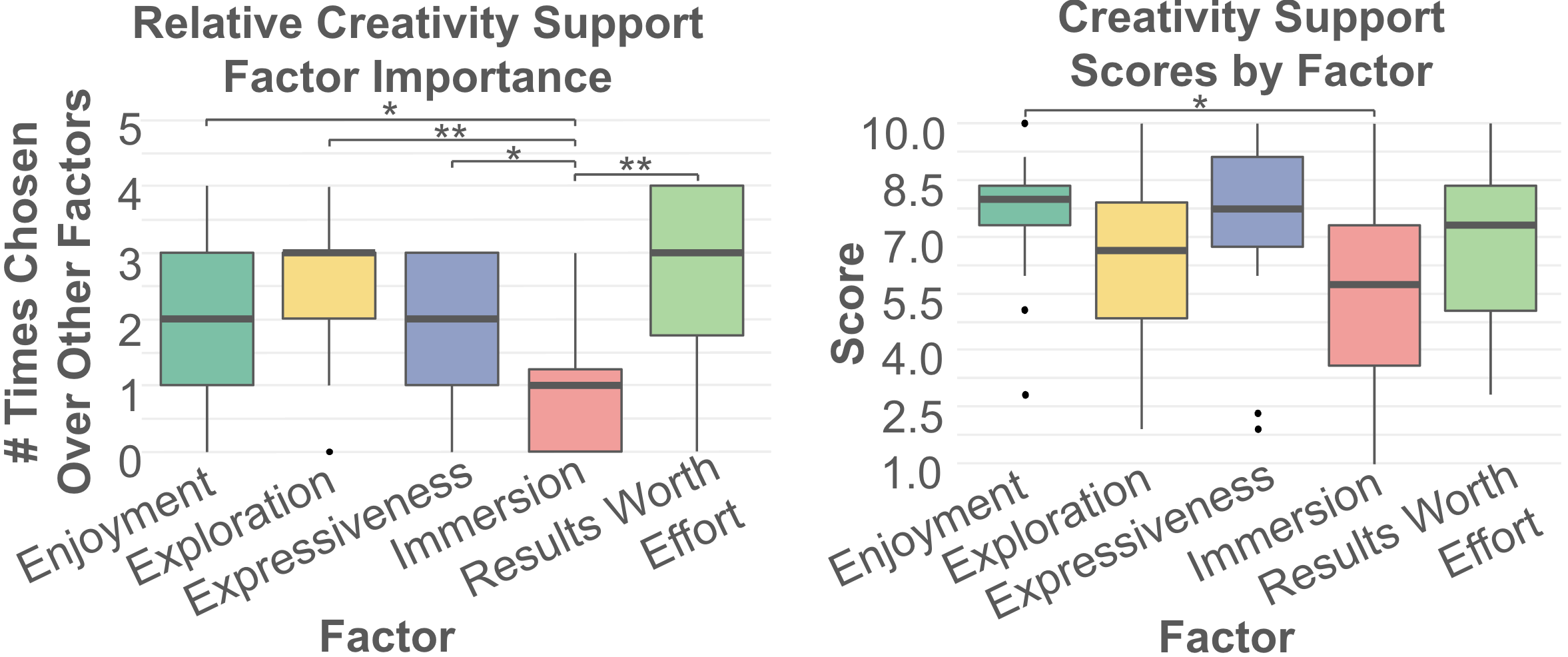}
    \caption{Left: Average number of times each CSI factor~\cite{Cherry2014} was selected as more important than another. Participants rated Results Worth Effort and Exploration as most important, with Immersion rated significantly lower than all other factors. Right: The scores participants gave \sysname{} by factor. Participants found \sysname{} enjoyable and expressive, but not necessarily immersive.}
    \label{fig:csi_factor_scores_and_importance}
\end{figure}

\textbf{Creativity Support.} \sysname{} received an average Creativity Support Index (CSI) score of 68.8 (SD = 18.0). CSI scores can be mapped to educational grading scales~\cite{Cherry2014}, and since our study was conducted in the UK, this corresponds to an `\textit{Upper Second-class Honours}'---the second-highest classification~\cite{panEuropeanGradingScales-karran}. Participants rated \textit{Results Worth Effort} (M = 2.65, SD = 1.50) and \textit{Exploration} (M = 2.50, SD = 1.24) as the most important factors in AR authoring. On a 1–10 scale, \sysname{} scored 6.65 (SD = 2.22) for Results Worth Effort and 6.36 (SD = 2.17) for Exploration. The highest-rated aspects of the system were \textit{Enjoyment} (M = 7.71, SD = 1.65) and \textit{Expressiveness} (M = 7.55, SD = 2.24). These results suggest participants valued the ability to explore and achieve meaningful outcomes---well-supported by \sysname{}---while also finding the experience engaging and expressive. See Figure~\ref{fig:csi_factor_scores_and_importance}.

\begin{figure*}[htbp]
  \centering
  \includegraphics[width=0.83\linewidth]{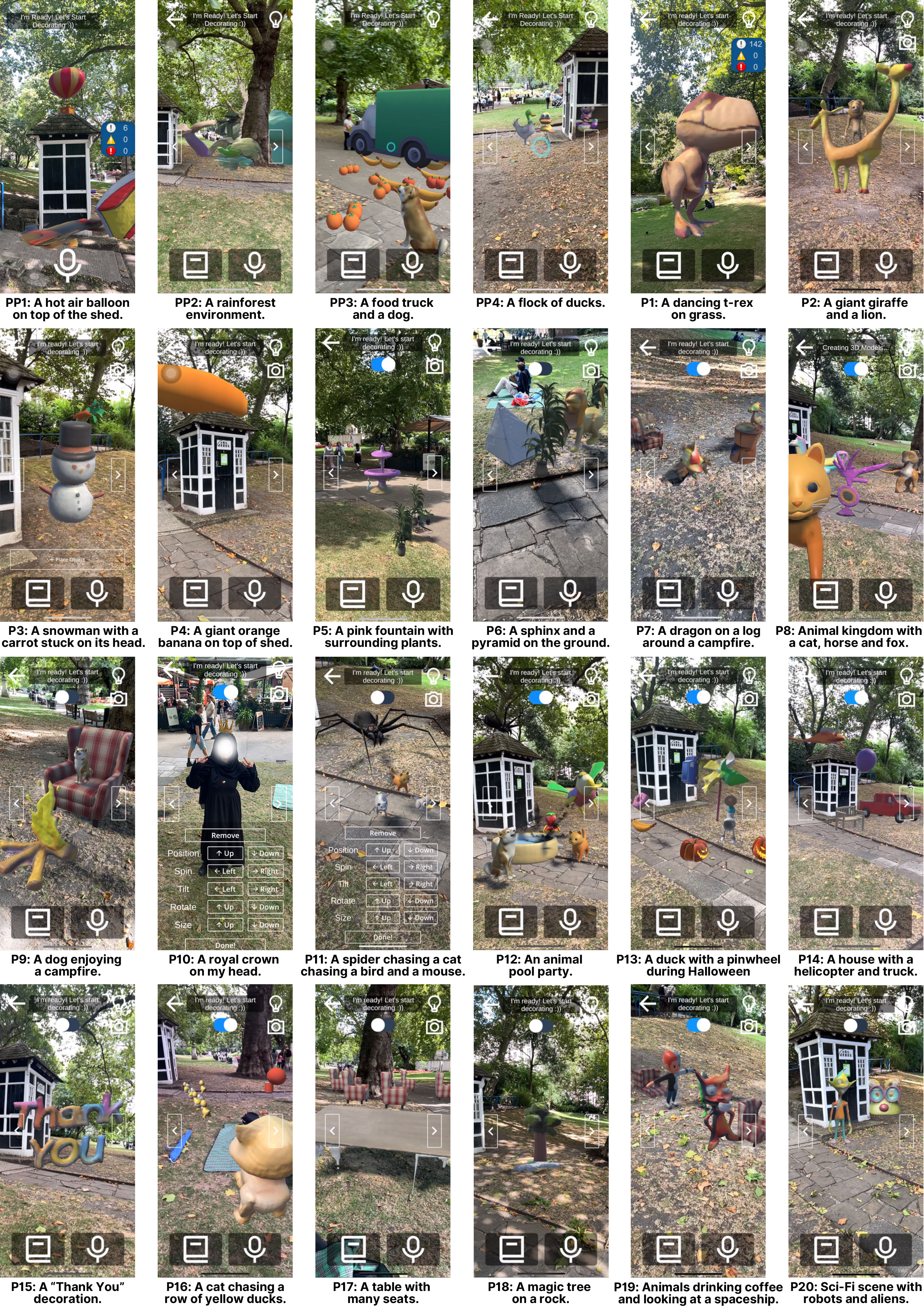}
  \caption{AR experiences created by participants (N$_{pilot}$=4; N$_{study}$=20) while interacting with the full \sysname{} prototype in Part 2. Users were encouraged to create freely without limitations. 
  `PP' denotes pilot participants and `P' study participants.}
  \label{fig:user_creations}
  \Description[AR experiences created by participants.]{AR experiences created by participants.}
\end{figure*}

\textbf{Participant Creations.} All participants successfully authored at least one AR scene. See Figure~\ref{fig:user_creations} for all 24 creations. These ranged from ``\textit{a sphinx and a pyramid rising from the ground}'' (P6) to ``\textit{a cat chasing a row of yellow ducks}'' (P16) and ``\textit{animals drinking coffee while watching a spaceship launch}'' (P19). Some built whimsical scenes for general audiences (7/20), while others designed for friends (5/20) or family (3/20). A few explored more story-driven experiences (4/20). Regardless of intent, 14 out of 20 participants explicitly mentioned having fun while using \sysname{}. The variety of creations suggests that \sysname{} effectively supported a wide range of authoring goals, demonstrating both flexibility and robustness in real-world use.

\textbf{Authoring Strategies with \sysname{}.} Most participants (18/20) preferred a mix of AI and manual tools. Typically, they began with AI-assisted mode to create a ``\textit{blueprint layout}'' (P5), followed by ``\textit{manually tuning the scene as needed}'' (P6). AI features were praised for enhancing creativity (20/20), flexibility (16/20), and expressiveness (3/20), though some found them ``\textit{too creative}'' (P5), leading to unexpected or undesired results (7/20). Others noted subpar asset quality (7/20) and slow generation times (3/20).

Manual tools were valued for their control, precision, and sense of ownership (19/20), as well as ease-of-use (7/20). However, manual editing was also seen as time-consuming and laborious (12/20), requiring ``\textit{physically moving and pressing many buttons}'' (P15). Some participants found selecting models from a preset list creatively limiting (4/20), while others struggled with tapping accuracy in busy environments due to ``\textit{fat finger}'' issues (3/20).

Two participants diverged from this hybrid workflow: P2 skipped manual mode entirely, describing AI outputs as ``\textit{fun and creative, even when inaccurate}'' and arrangements ``\textit{correct enough}''. P19 avoided AI tools altogether due to slow generation times. Yet when asked how they would ideally use \sysname{} once AI and manual modes improved, all 20 participants indicated they would prefer a mix of both. As P4 put it: ``\textit{AI helped me be more creative and quickly place objects. But even when it was right, I still wanted to tweak things manually. It felt more rewarding when I had the final say.}''

Similar to Part 1, participants preferred the freedom to use AI and manual tools as needed. For brainstorming, however, participants relied solely on the AI agent. Eleven found it helpful, particularly when stuck or unsure what to create next. They especially appreciated how the agent suggested ideas aligned with their environment or theme. P20, for instance, began with a vague Sci-Fi idea and found the AI helpful in ``\textit{refining my idea into something more specific and creative}'', which led to creating an alien and a robot. Still, several participants (7/20) wished the agent could do more---holding a back-and-forth conversation (5/20), asking clarifying questions (4/20), and eventually generating an entire scene once the idea was fully formed (6/20). As P7 reflected, ``\textit{The AI adds flexibility, but also demands that you know exactly what you want and how to describe it},'' pointing to the potential for more collaborative, guided brainstorming and authoring workflows.

\subsection{Brainstorming Future of \sysname{}}
In Part 3, participants shared ideas for improving \sysname{} and envisioned how they might use it in the future. Below, we synthesize limitations they identified and their proposed enhancements.

\textbf{AI Creativity.} While participants agreed the AI was generally more creative than they were, they differed on whether that creativity was actually beneficial. 13 participants appreciated the AI’s inventive and surprising results---P2 remarked, ``\textit{It gave me a humanoid lion and a two-headed giraffe... I love the randomness of it. I’m just excited to see what it will create next!}'' Others found the AI ``\textit{too creative}'' (P5), generating content that clashed with their intent. For instance, P5 requested a fountain and received a pink one---possibly because previous objects they had generated were pink---when they had envisioned a typical stone fountain: ``\textit{Creativity can be a double-edged sword.}'' P1 also raised concerns that an unmoderated AI could produce inaccurate or even inappropriate content, especially for children.

To manage AI creativity, participants proposed several ideas. P14 wanted the AI to clarify ambiguous requests through follow-up questions, rather than making assumptions: ``\textit{If I ask for a creature but don’t specify the color, the AI should ask, `do you want it yellow, purple, or something else?' We should talk back and forth until both of us are ready to build something.}'' P18 suggested a ``\textit{creativity slider}'' for more granular control over AI outputs. Participants also appreciated being able to choose from multiple AI-generated options, helping them ``\textit{ignore results that don’t fit}'' (P8).

\textbf{Creating Dynamic AR Scenes.} Many participants wanted their scenes to feel alive and reactive, not just static. They suggested adding animations (8/20), music (5/20), and event triggers (3/20). For example, P5 wanted a water fountain with flowing water, P17 imagined dogs running in circles, and P20 envisioned horror scenes with eerie sounds: ``\textit{That would make it more realistic, especially if the rendering quality is more like a cartoon.}'' Additionally, P8 hoped virtual creatures could respond to touch (\textit{e.g.,} a dog smiling when petted), while P14 suggested NPC-like interactions where virtual humans or animals could talk or bark back in a conversational manner. Still, P18 felt the current features ``\textit{cover the basics needed to create a simple AR scene},'' but hoped future improvements would focus on AI generation quality and speed.

\textbf{Sharing Creations.} 11 participants expressed interest in sharing their AR creations. Some preferred sharing photos or videos (P9, P12, P16), while others (P6, P7, P20) wanted to distribute full AR scenes for others to download and experience. P5, P7, and P19 proposed a searchable catalog of AI-generated models with user ratings: ``\textit{If I had a catalog, I could just type in `pink dolphin' and see what others have used. That would drive inspiration and save me time}'' (P5). P7 added that ratings could help users assess model quality before choosing. To further personalize shared assets, P3, P6, and P16 suggested allowing users to customize elements like color. Finally, P1 emphasized that public sharing could help enforce content safety and appropriateness.

\textbf{AI Explainability.} Nine participants wanted clearer explanations from the AI about its progress and actions. Currently implemented messages like ``\textit{Understanding Your Surroundings}'' and ``\textit{Creating 3D Models}'' were seen as too vague. As P4 explained, ``\textit{Instead of just ‘thinking’ or ‘processing,’ a more detailed explanation of what's been done would be nice, just so I know the AI heard me right, how much longer I have to wait, and what it will eventually do to my environment.}'' Participants also wanted better feedback during AI processing to know whether they could continue interacting, such as looking around or making manual changes. That said, P18 cautioned against overloading users with information, suggesting that even a brief log would help: ``\textit{Long messages will go unread. Just tell me what the AI heard and what it's doing.}''

\textbf{Access Barriers.} Participants also raised accessibility concerns regarding speech input. P14 and P18 noted misrecognition of non-standard accents (\textit{e.g.,} ``\textit{bowl}'' interpreted as ``\textit{ball}''), while P5 highlighted issues for users with speech impairments or in noisy environments: ``\textit{If kids are screaming in the background, it might be easier not to speak out loud.}'' While speech input was chosen for its naturalness, participants emphasized the importance of offering alternatives to ensure broader accessibility.

\textbf{Envisioning Future Use Cases.} When asked where and how they might use \sysname{} in the future, participants proposed a wide range of scenarios. Popular ideas included designing mini or board games (P9, P13, P14, P15) and transforming mundane environments---such as turning lecture halls into botanical gardens or adding a beach to an office (P5, P7, P8). Some envisioned practical uses like visualizing furniture layouts (P2, P17) or using AR pets for stress relief (P1, P3). Others imagined playful experiences, such as hiding AR Easter eggs for friends (P4, P11) or creating immersive horror games (P7, P20). P10 even envisioned placing themselves inside the scene: ``\textit{I want to wear a crown, sit on a throne in the middle of a desert, and be surrounded by flowers.}'' Overall, participants were excited to use \sysname{} anywhere---from their homes (P2, P5) to parks (P5) and outdoor landmarks (P19).

\section{Discussion}
\sysname{} combines outdoor scene understanding, fast 3D asset generation, and LLM-driven speech interactions to advance AI-assisted AR authoring. Our study revealed that users often began with AI to generate a creative scene blueprint, then refined it manually for greater control---enabling diverse, \rev{accurate,} and expressive creations. Here, we provide suggestions for AI-assisted AR authoring tool designs, discuss the broader implications of AI creativity and assistance, and outline limitations and future directions.

\begin{figure*}[htbp]
    \centering
    \includegraphics[width=0.9\textwidth]{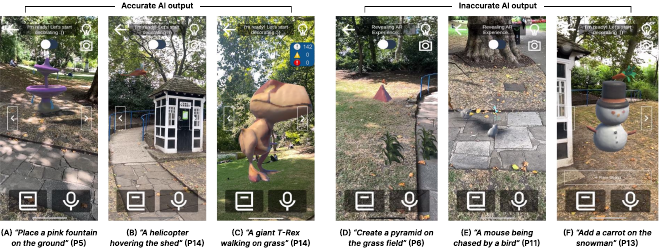}
    \caption{Examples of accurate and inaccurate AI-generated scene blueprints before any manual input. (A–C) show scenes that align with user intent across object type, placement, and orientation, though users may have made later edits. (D–F) illustrate common issues: clipped geometry (D), incorrect facing direction (E), and imprecise part-level placement (F).}
    \label{fig:ai-generations}
    \Description[Examples of accurate and inaccurate AI-generated scenes prior to any manual input.]{Examples of accurate and inaccurate AI-generated scenes prior to any manual input.}
\end{figure*}

\subsection{Design Implications for AR Authoring Tools}
Throughout the study, participants indicated preferences for AI use and proposed a wide range of improvements and future features for \sysname{}. We summarize and expand on these suggestions.

\textbf{What Role Should AI Play in AR Authoring Workflows?} Our key takeaway is that users expect a blend of AI-assisted and manual tools when authoring AR environments---they want to co-create with AI, not just rely on it. While AI offers creativity and expressivity, manual tools provide the control needed to fine-tune scenes and feel ownership over the result. All but two participants combined both during free-form authoring: they reviewed AI-generated blueprints, then refined one to better match their creative intent. AI sped up early prototyping, helping users bring ideas to life with less active input and decision-making, while manual adjustments enabled greater precision and reduced frustration by offering a way to correct AI errors. We recommend that future iterations of \sysname{} continue supporting hybrid workflows, consistent with human-AI design guidelines~\cite{mixedInitiativeUserInterfaces-horvitz,humanAiGuidelines-amershi}. Ultimately, users seek outcomes that justify their effort---AR scenes that best reflect their imagination---which often requires both the creative freedom of generative AI and the precision of manual control.

\textbf{How Much AI Creativity is Too Much?} 
AI’s creativity can be a double-edged sword---both engaging and frustrating. Some participants enjoyed the AI’s playful interpretations---like P2’s whimsical two-headed giraffe---while others felt such outputs strayed too far from their intent. This tension suggests \sysname{} should avoid extremes: being too rigid, where the AI follows only literal instructions, or too free, where it produces imaginative but irrelevant content. Following the \textit{Human-Centered Artificial Intelligence (HCAI)} framework~\cite{hcai-shneiderman}, we recommend giving users ways to adjust AI creativity (\textit{e.g.,} a ``\textit{creativity slider}'' akin to an LLM’s temperature setting) while supporting rapid iteration so users stay in control.

\textbf{What Might Future AR Authoring Look Like?} 
Our findings point to a future AR authoring workflow where users and AI co-create through iterative conversation, refining ideas together until both ``agree'' on what to build. Once aligned, the system could generate a full scene blueprint. For example, a user might say, ``\textit{Turn this playground into a coral reef}'', imagining an experience where kids can explore and learn about marine life. The AI might suggest creative details like, ``\textit{Let’s add a surgeonfish and a parrotfish, since they’re commonly found in coral reefs}'', or ask clarifying questions such as, ``\textit{What color corals would you like?}'' rather than making assumptions. This kind of dialogue lets users guide the creative direction without needing to specify every object or detail---alleviating the burden of constant input and decision-making. The same workflow can extend beyond 3D assets to include music (\textit{e.g.,} sea breeze), animations (\textit{e.g.,} fish flapping their fins), and event triggers (\textit{e.g.,} picking up corals that break). Once both parties feel ready, AI agents can build the scene. \rev{To support this, AR authoring tools need scene understanding to position, rotate, and scale multiple objects appropriately---reducing the user's workload of arranging each asset manually.} Users could then make quick manual edits to fine-tune the result, and ideally, share it with others. While \sysname{} already supports conversational brainstorming, real-time full-scene generation remains limited by current technology: even our fast asset generation pipeline takes 20–30 seconds per model, making scene-level creation too slow for interactive use. This vision also aligns with prior work like LLMR's Planner agent~\cite{llmr-delatorre}, which also supports collaborative scene ideation---but primarily targets VR and still struggles to generate complete scenes efficiently. However, as generative models continue to improve, conversational AR authoring at scale may soon be possible.

\subsection{Challenges in AI-powered AR Authoring}
This work contributes to both HCI and computer vision by integrating outdoor scene understanding and fast 3D asset generation into a simple, speech-driven system for AI-assisted AR authoring. However, our study revealed limitations that impacted user experience. For example, scene understanding sometimes lacked granularity, leading to visual misalignments, while asset generation, though significantly faster than prior work, still required around half a minute---affecting perceived usability. Below, we reflect on key technical challenges. We also dig deeper into CV–specific challenges in Sections 1–3 of the Supplementary Materials, including depth map enhancement, scene understanding, and 3D asset generation.

\textbf{Scene Understanding Accuracy and Granularity.} We represent real-world objects as 3D bounding boxes to keep the scene graph compact and make spatial reasoning easier for LLMs. However, this abstraction can limit precision in AI-generated scene blueprints. For example, a sloped ground in our study environment was enclosed in a tall bounding box. \rev{When users asked for virtual objects to be placed on this surface, aligning to the bounding box’s maximum y-value caused them to float near the bottom of the slope, while using the minimum y-value led to clipping near the top.} Placement on irregular, multi-part shapes like the pig statue in Figure~\ref{fig:proof_by_demonstration}A was also challenging. A hat worked reasonably well by aligning its base to the statue’s bounding box top, \rev{but clothing---intended for the body---was} harder to position due to the lack of part segmentation. Even with the hat, minor misalignments occurred because the statue’s ears extended above its head, meaning the maximum y-value did not match the intended placement point. Figure~\ref{fig:ai-generations} illustrates scenes created solely by AI, including both successful and unsuccessful examples. While 3D bounding boxes offer an efficient abstraction, future work should explore richer representations to support more precise interactions\rev{---such as directly leveraging depth maps~\cite{Du2020} or point clouds~\cite{wang2023pointshopar}---though these formats are less readily compatible with LLM-based pipelines compared to textual scene graphs.}

\textbf{Speed and Quality of Asset Generation.} In-situ AR authoring demands fast, high-quality 3D mesh creation. Although our pipeline generated assets faster than prior work with minimal sacrifice of quality, participants still found the 30 second wait disruptive. Asset quality was also occasionally lacking: some models were flat (princess in Figure~\ref{fig:proof_by_demonstration}B), incomplete (knight missing legs in Figure~\ref{fig:proof_by_demonstration}B), had holes (castle in Figure~\ref{fig:proof_by_demonstration}B), or lacked detail (Mayan person in Figure~\ref{fig:proof_by_demonstration}C). Interestingly, some errors were viewed positively---P2 described a two-headed giraffe as ``\textit{fun and exciting}''---but overall, more reliable asset quality is needed. At the time of development, we used InstantMesh~\cite{xu2024instantmesh}, then state-of-the-art, but more capable models like \textit{LATTE3D}~\cite{Xie2024}\rev{, \textit{TRELLIS}~\cite{xiang2025trellis}, and \textit{Hunyuan3D}~\cite{zhao2025hunyuan3d20}} are emerging. Since \sysname{} is modular, these can be easily integrated.

\rev{\textbf{Beyond Static Scenes.} While \sysname{} supports AR authoring across a wide range of static outdoor scenes, dynamic content remains an open challenge. Our scene understanding pipeline relies on pre-scans to reduce user burden and enable more complete spatial understanding. We use scene graphs for their compact, textual structure that LLMs can reason over---but they do not reflect real-time changes, such as a moved bench or a person walking through the scene. As a result, while we can place a hat on a statue, we cannot place it on a moving person. Authoring truly dynamic scenes would also require richer support for animation, sound, and interactivity. Audio could be integrated using generative models~\cite{sonifyar-su} (\textit{e.g.,} \textit{AudioLDM}~\cite{liu2023audioldm}, \textit{MusicLM}~\cite{agostinelli2023musiclm}), and triggered events could build on prior systems that generate code~\cite{llmr-delatorre, dreamCodeVR-giunchi, llmer-chen}. Animation, however, is particularly challenging: most prior work uses simple scripted motions~\cite{llmr-delatorre, dreamCodeVR-giunchi, llmer-chen} or assumes pre-rigged assets~\cite{huang2024realtimeanimationgeneration}, which is incompatible with our use of generated 3D models. Auto-rigging remains unreliable, and low-quality animation risks breaking immersion. Therefore, we chose to study animation needs in AR authoring qualitatively (\textit{e.g.,} P5: ``\textit{flowing water}''). Future work should explore how to incorporate dynamic changes and behaviors into AR authoring~\cite{srinidhi2024xair} to further enhance creative flexibility.}

\textbf{System Latency.} Latency remains a core challenge for in-situ AR systems---users expect responsiveness and may find even sub-minute delays disruptive, especially outdoors. While \sysname{} achieves significantly faster runtimes than prior systems (\textit{i.e.,} $33.92\pm5.83$ seconds), current speeds can still interrupt the flow of in-situ authoring. Because true real-time performance remains difficult to achieve, future tools should offer meaningful feedback (\textit{e.g.,} progress indicators, estimated wait times) and support multitasking---such as manually editing objects while waiting for AI responses. As generative models improve, latency will likely decrease, though offloading to remote servers may remain necessary given the limited computational power of today’s AR devices.

\subsection{Limitations \& Future Directions}
\rev{This work has several limitations.} First, we did not support multi-user co-creation. Several participants expressed interest in sharing or building scenes together, suggesting opportunities to study collaborative AR authoring~\cite{numan2024spaceblender}. Second, our user study was limited to a single location. While our technical evaluation shows that \sysname{} can generalize to diverse outdoor settings, future studies should explore a broader range of environments (and perhaps with other demographics, such as children). \rev{Third, while \sysname{} currently runs on phones, future work could explore deploying it on AR headsets, which may enable new interactions but also raise challenges around social acceptability and physical comfort during extended public use.} Fourth, as discussed earlier, improving scene understanding, asset quality, system latency, \rev{and support for dynamic scene authoring} remains important. Future scene understanding pipelines should also be evaluated on larger outdoor datasets. Finally, although \sysname{} depends on precomputed scene graphs, participants did not perform scanning themselves. While the system is designed to scale with existing large-scale point cloud datasets, future work could examine how users scan scenes and how systems might better support that process~\cite{Brummelen2024}.

\section{Conclusion}
We present \sysname{}, a novel system that advances AI-assisted AR authoring through outdoor scene understanding, fast 3D asset generation, and LLM-driven speech interactions. Our technical evaluation and user study show that users can create diverse AR scenes in different real-world settings. Challenges remain---including improving scene understanding granularity, asset quality, latency, and collaborative AI support---but this work takes a step toward making personalized AR authoring as simple as speaking your imagination.

\bibliographystyle{ACM-Reference-Format}
\bibliography{paper}

\end{document}


\title{Supplementary Materials for \textit{\sysname{}: AI-Assisted In-Situ Authoring in Augmented Reality}}

\maketitle

\tableofcontents

\section{Depth Map Enhancement Using Monocular Depth Estimator}

\begin{figure*}[!h]
    \centering
    \includegraphics[width=0.95\textwidth]{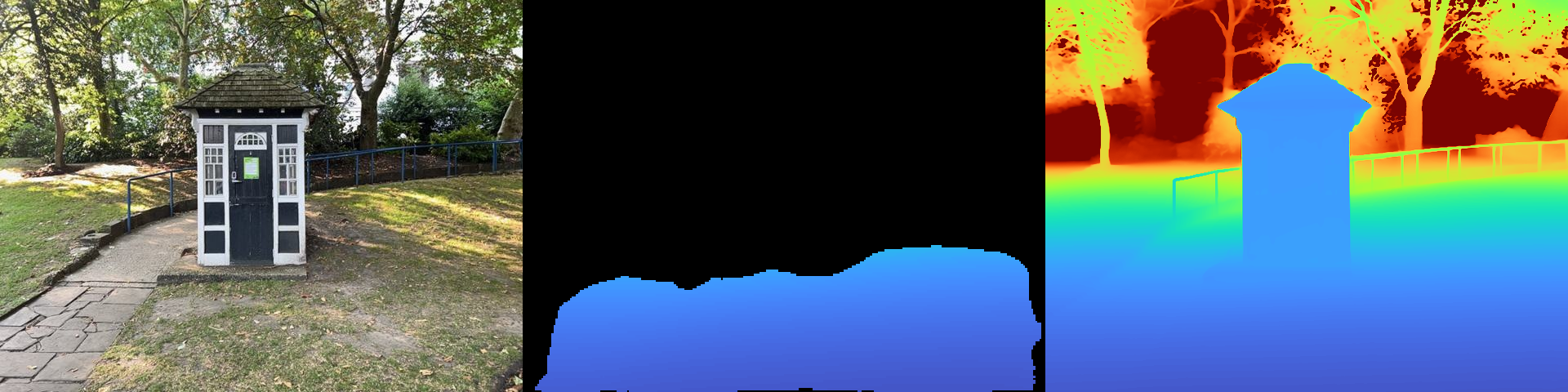}
    \caption{Comparing inputs and output for monocular depth enhancement: The figure shows the source image (left), the sensor's original depth map (center), and the improved metric monocular depth map after enhancement (right). In the depth maps, red colors signify points farther from the viewer. The depth maps use a colormap where red represents farther points.}
    \label{fig:depth_completion}
\end{figure*}

Figure \ref{fig:depth_completion} shows an example. To overcome sensor inaccuracies (especially outdoors for objects farther than 6-10m), we scale a monocular estimator's output \cite{yang2024depth_anything_v2} using valid sensor depth points, yielding dense metric maps.

\section{Scene Understanding Pipeline: Limitations and Future Works}

\begin{figure*}[!h]
    \centering
    \includegraphics[width=0.48\textwidth]{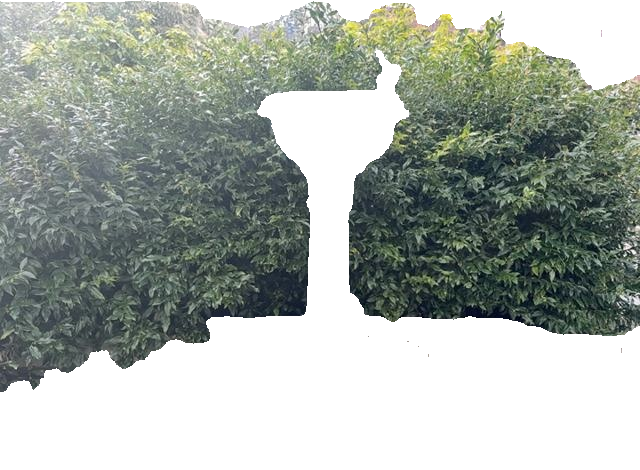} 
    \hfill
    \includegraphics[width=0.48\textwidth]{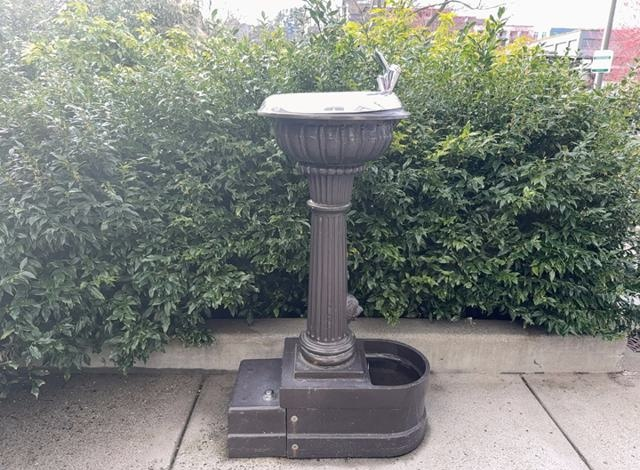} %
    \caption{Object and Context crops for a failure case. In this case, the VLM tried to \textit{look through} the white mask and predicted \textrm{drinking fountain} instead of \textrm{bush} or \textrm{vegetation}.}
    \label{fig:classification-error}
\end{figure*}

The proposed scene understanding pipeline is effective in generating compact scene graph with semantic labels. However, we have identified three main issues that can be addressed in future work. First, results generated by our method are conditioned by the initial set of masks predicted by OpenMask3D \cite{takmaz2023openmask3d}. This method, trained indoors \cite{dai2017scannet}, detects masks in outdoor scenarios, but suffers from a domain gap. Nevertheless, this space is moving rapidly, and more robust methods were released after our user study. For instance, the method in \cite{Huang2023Segment3D}, which has an example of scene understanding in an outdoor scenario, may be able to improve our results. A better initial set of masks is crucial to remove the preliminary filtering step: in fact, this filter acts without prior scene information and potentially removes or merges valid masks. Second, our clustering refinement is directly affected by the quality and the consistency of the labels predicted by the vision-language model (VLM). When these labels are incorrect, the results of the clustering step on top of the semantic point cloud can cause errors. For example, this may happen when the VLM ignores the prompt and tries to classify what has been covered by the mask, as shown in Figure \ref{fig:classification-error}. In this case, the misclassified bounding box---labeled as \textsf{drinking fountain} instead of \textsf{bush}---increases the chances of it being merged with the bounding box of the actual fountain, thereby generating a noisy bounding box for the fountain. In our experiments, we have noticed that the majority of the predicted labels are stable across different runs; however, a few of them might vary. This effect explains the slightly higher variation observed in the last row of the benchmark (see Table 2 of the main document). Validation checks could be enforced in future work to mitigate the problem. Third, our scene graphs are snapshots of the world at the time of scanning. This is generally not a problem because they mainly represent static objects (dynamic objects---such as parked cars---could potentially be removed using semantic labels), however we do not include real-time scene understanding components during the user experience. Future work can address this limitation by incorporating live components to further enhance the scene graphs.

\section{AI generated 3D Assets}
As reported in the main paper, the \textit{Action Plan} Agent can invoke the generation of a virtual 3D asset on-the-fly. 

\begin{figure*}[htbp]
    \centering
    \begin{subfigure}{0.48\textwidth} %
        \includegraphics[width=\textwidth]{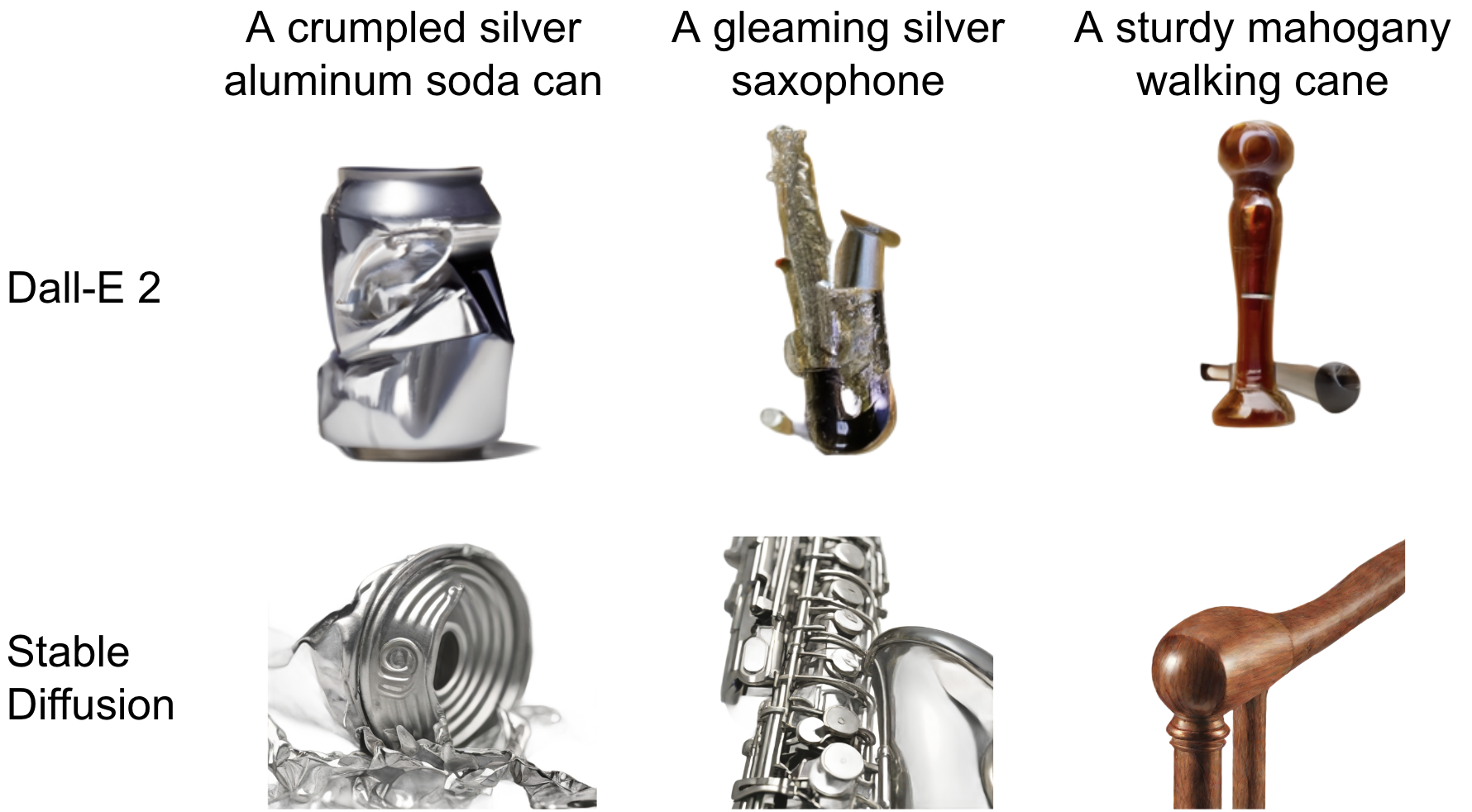}
        \caption{Generated images.}
        \label{fig:subfig1}
    \end{subfigure}
    \hfill %
    \begin{subfigure}{0.48\textwidth} %
        \includegraphics[width=\textwidth]{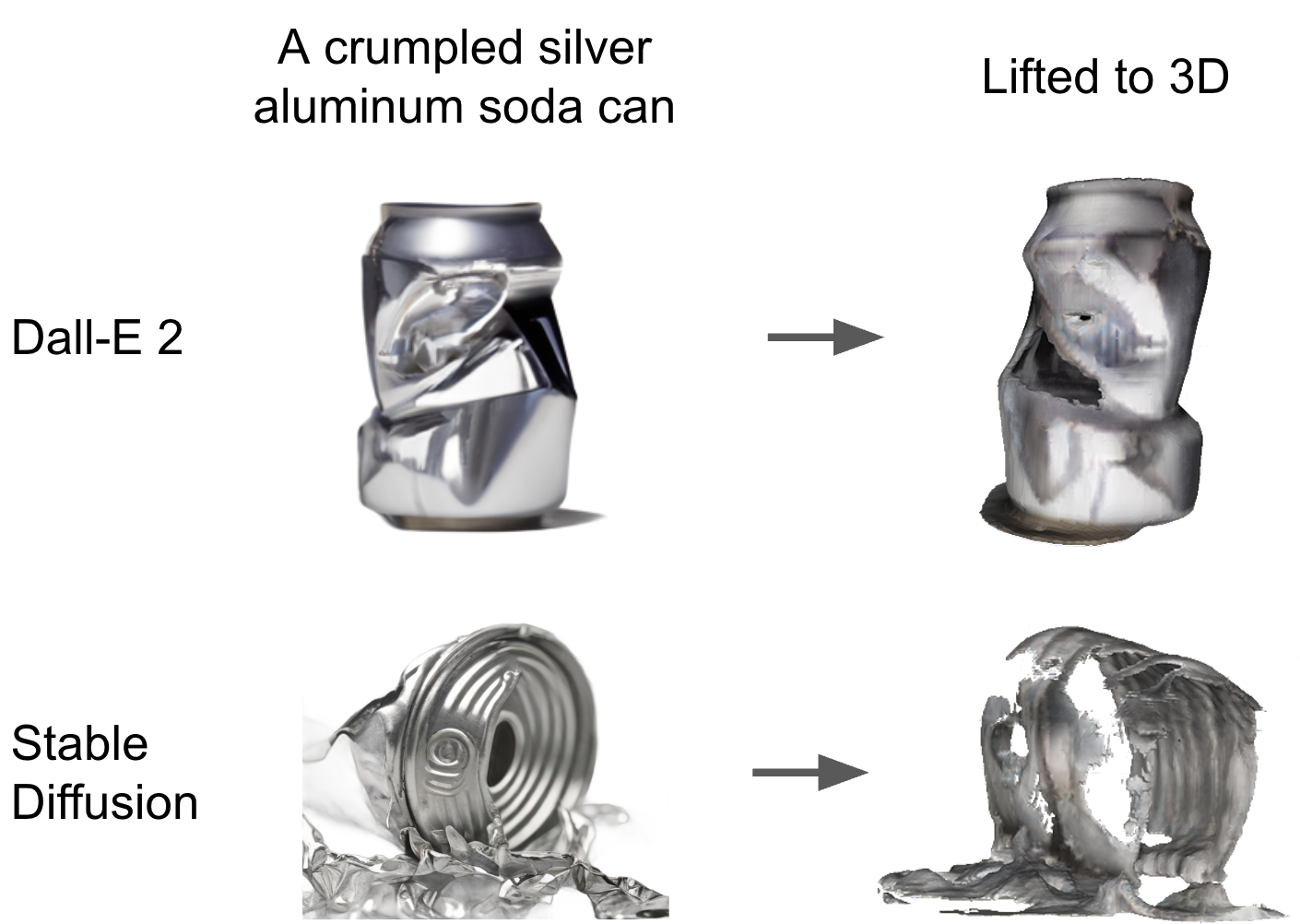}
        \caption{Generated 3D assets}
        \label{fig:subfig2}
    \end{subfigure}
    \caption{Comparison between DALL-E2 \cite{ramesh2022hierarchical} and StableDiffusion Turbo \cite{sauer2023adversarial}. In these examples the inpainting strategy helps in generating fully visible objects. In contrast, StableDiffusion Turbo generates high-quality images of partially visible objects. When lifted in 3D, partially visible assets can lead to ambiguous or poorly defined 3D models. We used prompt-boosting for both the strategies.}
    \label{fig:partially-visible-assets}
\end{figure*}

We chose DALL-E2 \cite{ramesh2022hierarchical} for its ability to generate images conditioned on input masks. Although its visual quality is inferior to other generative models, such as Stable Diffusion \cite{rombach2022latent}, the inpainting constraint often yields more complete objects on a plain background, which directly facilitates background removal \cite{qin2022} and image-to-3D reconstruction \cite{xu2024instantmesh}. Figure \ref{fig:partially-visible-assets} shows examples of assets generated by StableDiffusion Turbo \cite{sauer2023adversarial} that are partially visible in the image. These assets can lead to ambiguous 3D models when lifted by InstantMesh \cite{xu2024instantmesh}. 

\begin{figure*}
    \centering
    \begin{minipage}{0.3\textwidth}
        \centering \textbf{Ours w/o prompt boosting}
    \end{minipage}
    \hfill
    \begin{minipage}{0.3\textwidth}
        \centering \textbf{Ours w/ prompt boosting}
    \end{minipage} \\ %

    \subfloat{\includegraphics[width=0.3\textwidth,trim=3cm 3cm 3cm 3cm, clip]{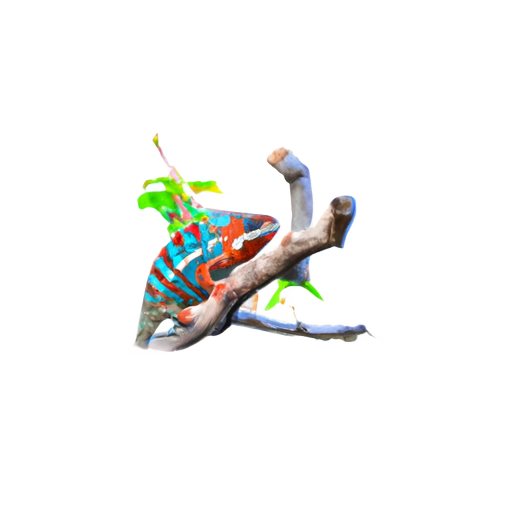}}
    \hfill
    \subfloat{\includegraphics[width=0.3\textwidth,trim=3cm 3cm 3cm 3cm, clip]{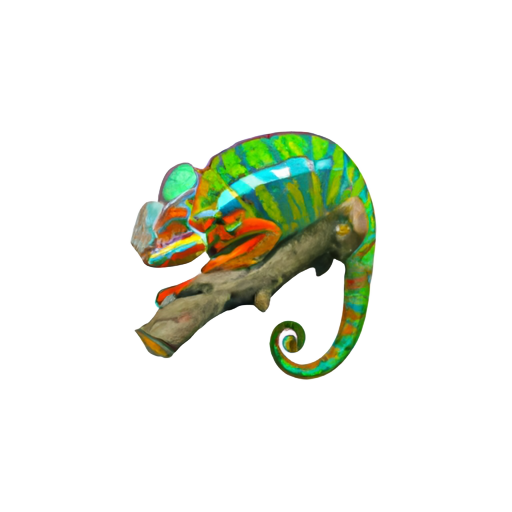}}

    \begin{minipage}{\textwidth}
        \centering \textit{A chameleon perched on a tree branch}
    \end{minipage} \\
    
    \subfloat{\includegraphics[width=0.3\textwidth,trim=3cm 3cm 3cm 3cm, clip]{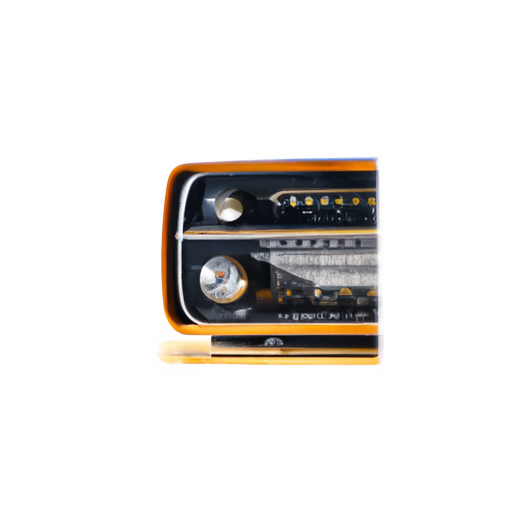}}
    \hfill
    \subfloat{\includegraphics[width=0.3\textwidth,trim=3cm 3cm 3cm 3cm, clip]{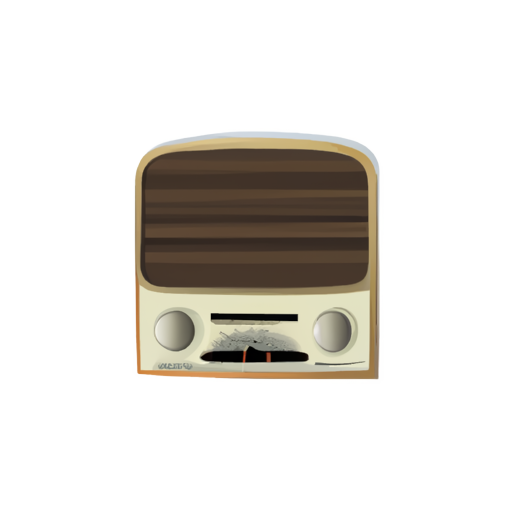}}
    
    \begin{minipage}{\textwidth}
        \centering \textit{A classic leatherette radio with dials}
    \end{minipage} \\
    
    \subfloat{\includegraphics[width=0.3\textwidth,trim=3cm 3cm 3cm 3cm, clip]{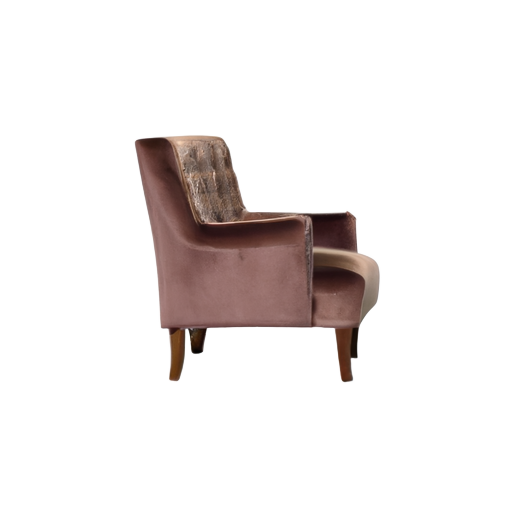}}
    \hfill
    \subfloat{\includegraphics[width=0.3\textwidth,trim=3cm 3cm 3cm 3cm, clip]{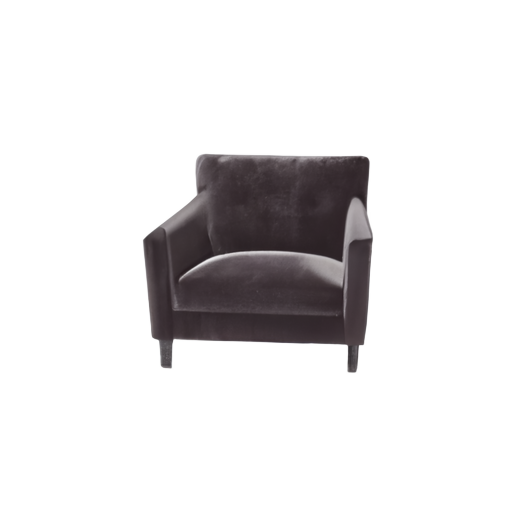}}

    \begin{minipage}{\textwidth}
        \centering \textit{A plush velvet armchair}
    \end{minipage} \\

    \caption{A qualitative comparison of generated images: without (first column) and with (second column) prompt boosting. Prompt boosting results in objects that are more fully visible.}

    \vspace{50px}
    \label{fig:prompt-boosting-comparison}
\end{figure*}

Figure \ref{fig:prompt-boosting-comparison} compares the images generated by our pipeline for the same prompts with and without using prompt boosting. Using prompt boosting results in assets that are better suited for 3D reconstruction. 

\section{Prompts}
In this section, we report the prompts used by our AI agents.

\begin{figure*}[htbp]
    \centering
    \includegraphics[clip,width=0.72\textwidth,trim=0cm 2.7cm 0cm 0cm]{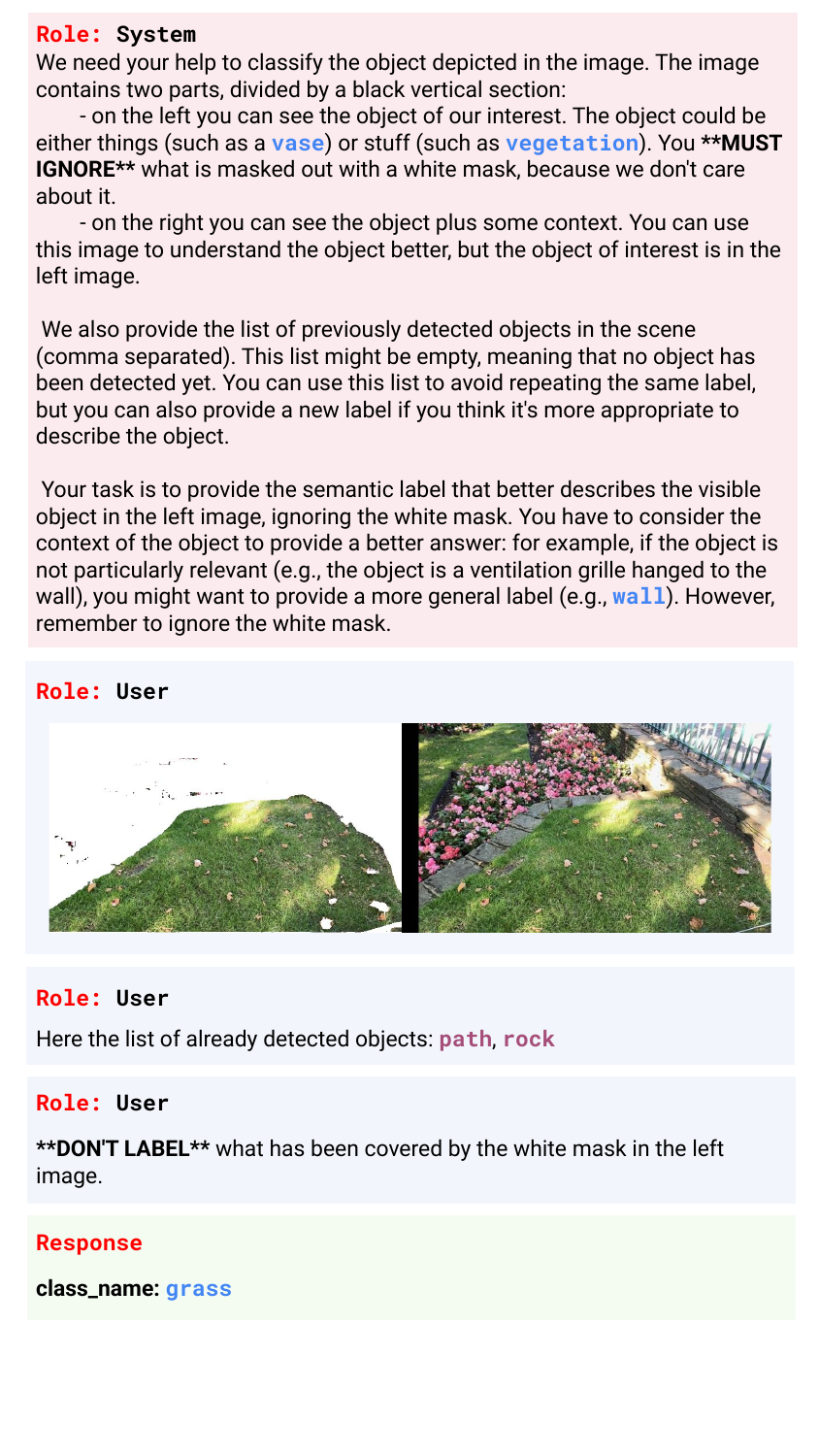}
    \caption{Example of the prompt used by the Object Classifier agent}
    \label{fig:classify_prompt}
    \Description{Example of the prompt used by the Object Classifier agent}
\end{figure*}

Figure \ref{fig:classify_prompt} presents an example of a prompt used by the \textit{Object Classifier} agent. Instructions are given to the agent in GPT's \textit{System mode} \cite{systemUserModesChatGPT}, while additional inputs---the object and context crops, as well as the list of previously predicted labels---are passed in GPT's \textit{User mode} \cite{systemUserModesChatGPT}. Finally, the agent replies with the semantic label of the object. It is worth noting that all inputs are generated by the system itself during execution, with no human intervention required. 

\begin{figure*}[htbp]
    \centering
\includegraphics[clip,,width=0.7\textwidth]{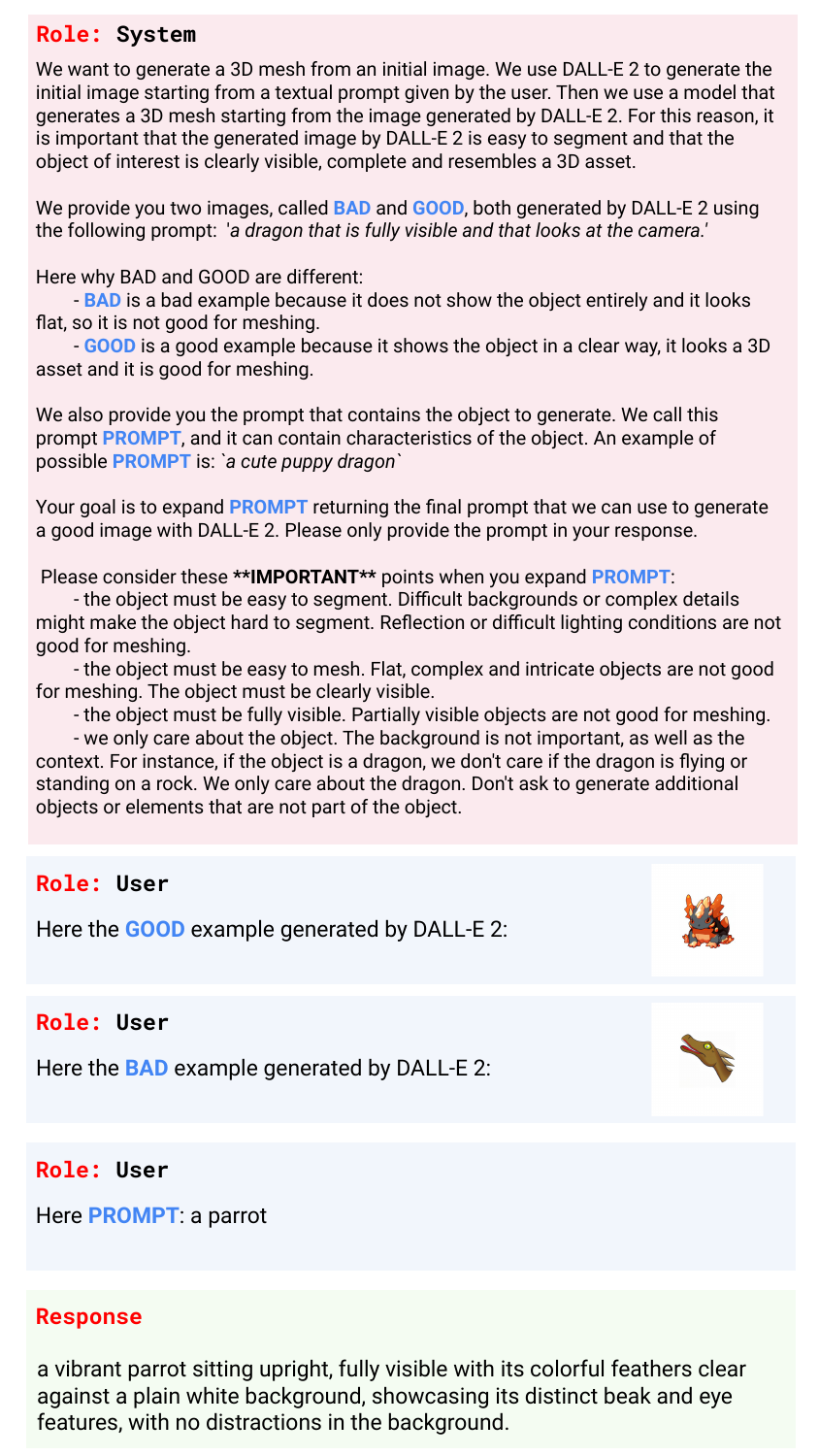}
    \caption{Example of the prompt used for \textit{Prompt Boosting} when generating new assets.}
     \Description{Example of the prompt used for \textit{Prompt Boosting} when generating new assets.}
    \label{fig:boosting_prompt}
\end{figure*}

Figure \ref{fig:boosting_prompt} shows an example of the prompt used by the \textit{Prompt Boosting} agent. Again, we use \textit{System mode} to instruct the agent about the task, and in \textit{User mode}, we provide two visual examples (these examples are not request-specific; we use the same example in every request) and the initial user prompt to boost. The agent replies with the boosted prompt, which is then used to generate an image suited for image-to-3D methods.

\begin{figure*}[htbp]
    \centering
\includegraphics[clip,width=0.9\textwidth]{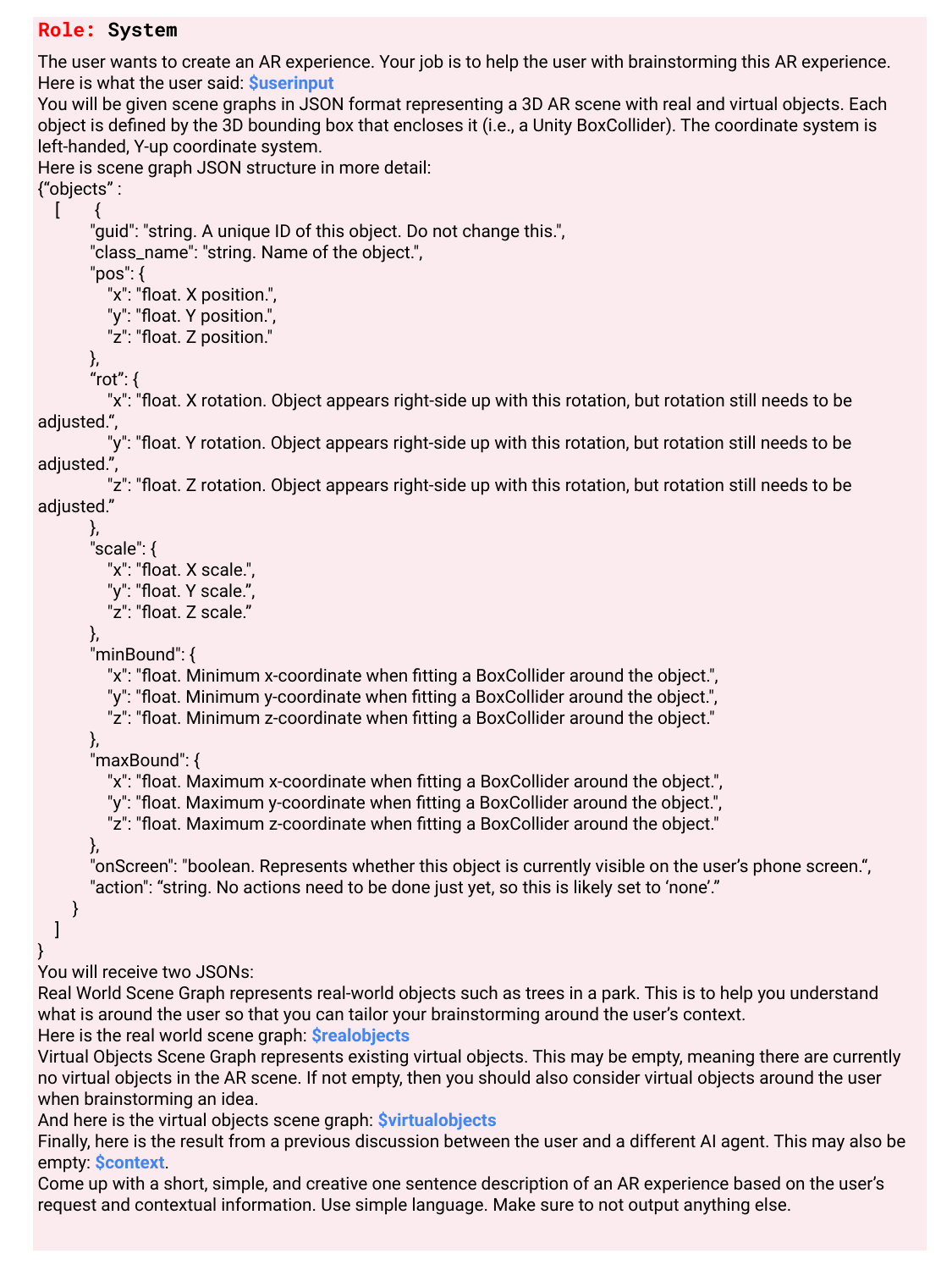}
    \caption{Prompt used by the \textit{Brainstorming} agent.}
     \Description{Prompt used by the Brainstorming agent.}
    \label{fig:brainstorm_prompt}
\end{figure*}

Figure \ref{fig:brainstorm_prompt} illustrates the prompt used by the \textit{Brainstorming} agent. This agent generates a short story-based description of an AR experience given a list of real and virtual objects in the scene, as well as conversations with the user.

\begin{figure*}[htbp]
    \centering
\includegraphics[width=0.9\textwidth]{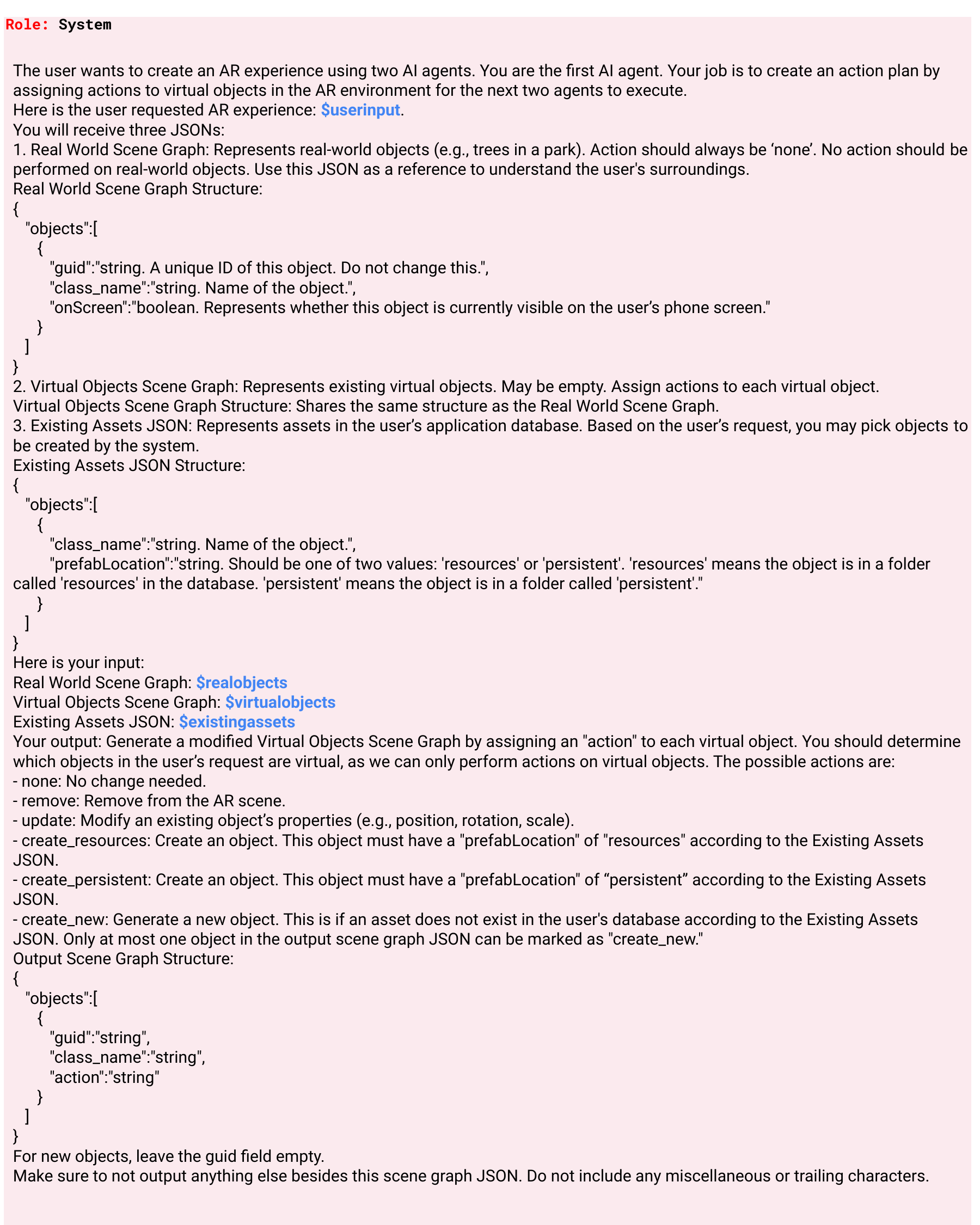}
    \caption{Prompt used by the \textit{Action Plan} agent.}
     \Description{Prompt used by the Action Plan agent.}
    \label{fig:plan_agent}
\end{figure*}

Figure \ref{fig:plan_agent} reports the prompt used by the \textit{Action Plan} agent. Based on the current state of the AR experience and a list of already-generated assets, the agent defines the \textsf{action} to apply to each existing and new virtual object to better align it with the user-described experience.

\begin{figure*}[htbp]
    \centering
\includegraphics[width=0.9\textwidth]{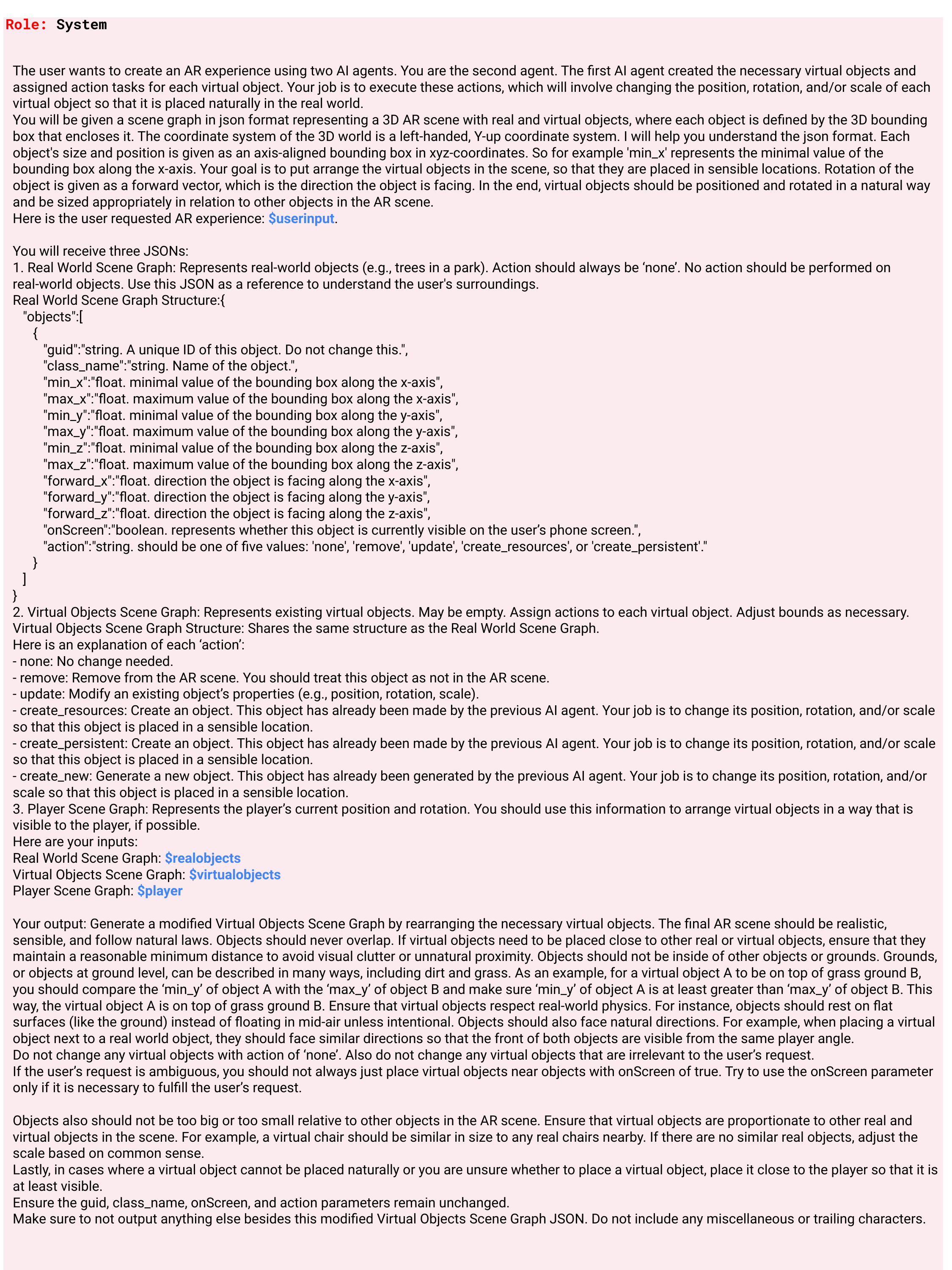}
    \caption{Prompt used by the \textit{Assembly} agent.}
     \Description{Prompt used by the Assembly agent.}
    \label{fig:assembly_agent}
\end{figure*}

Finally, Figure \ref{fig:assembly_agent} shows the prompt adopted by the \textit{Assembly} agent. This agent receives the actions planned by the \textit{Action Plan} agent (\eg, modifying an object's position) and applies them to virtual objects.

\vspace{30px}

\section{Study Materials}
This section contains references to the validated questionnaires we used, as well as custom questions and semi-structured interview questions we developed for the study.

\subsection{Pre-Study Questionnaire}
The pre-study questionnaire asked participants various demographics and experience questions, as shown in Listing~\ref{lst:pre-study-survey}.

\begin{lstlisting}[caption={The pre-study questionnaire.},label={lst:pre-study-survey},frame=single]
1. What is your age?
[Number]

2. What gender do you self-identify with? (E.g. woman, non-binary, man, etc.)
[Short answer]

~ Prior Experience with Technology ~
3. How familiar are you with augmented reality (AR), such as Pokemon GO, Apple Vision Pro, Meta Quest, etc.?
[Options: Not at all familiar to Very familiar on a 5-point scale]

4. List any AR technologies and applications you have used before and what you used them for (or write "None").
[Long answer]

5. How often do you use augmented reality (AR) technologies or applications?
[Options: Multiple times each day to Never on a 7-point scale]

6. How familiar are you with artificial intelligence (AI) chat systems, such as chatbots, ChatGPT, etc.?
[Options: Not at all familiar to Very familiar on a 5-point scale]

7. List any AI chat systems you have used before and what you used them for (or write "None").
[Long answer]

8. How often do you use AI chat system(s)?
[Options: Multiple times each day to Never on a 7-point scale]

9. How familiar are you with 2D creativity tools, such as the Adobe suite (e.g. Photoshop, Premiere Pro), Figma, etc.?
[Options: Not at all familiar to Very familiar on a 5-point scale]

10. List any 2D creativity tools you have used before and what you used them for (or write "None").
[Long answer]

11. How often do you use 2D creativity tool(s)?
[Options: Multiple times each day to Never on a 7-point scale]

12. How familiar are you with 3D creativity tools, such as Maya, Blender, Unity, Unreal, etc.?
[Options: Not at all familiar to Very familiar on a 5-point scale]

13. List any 3D creativity tools you have used before and what you used them for (or write "None").
[Long answer]

14. How often do you use 3D creativity tool(s)?
[Options: Multiple times each day to Never on a 7-point scale]

\end{lstlisting}

\subsection{Part 1: Comparison Task Questionnaires}
There were three comparison task questionnaires in the first part of the study. We provided the first one to participants after every trial of the five features (\eg, brainstorming, modifying objects, etc.) for one of the three systems (A, B or C). This meant each participant completed this questionnaire 15 times (5 features by 3 systems). It is shown in Listing~\ref{lst:per-feature-comparison-survey}. This survey included custom questions about creativity, the UMUX-LITE \cite{Lewis2013} questionnaire to assess usability, and questions from the NASA-TLX \cite{Hart2006} to assess task load.

The second comparison task questionnaire asked the participants which system they preferred overall and why, as shown in Listing~\ref{lst:all-feature-comparison-survey}. The final comparison task questionnaire provided participants with descriptions of how new features (\eg, adding music, animating objects, etc.) would work if they were added to System A, B or C, and asked which version participants would prefer and why. This is shown in Listing~\ref{lst:new-feature-comparison-survey}.

\begin{lstlisting}[caption={The first comparison task questionnaire, which participants completed after every trial of a new system feature.},label={lst:per-feature-comparison-survey},frame=single]
1. Which feature did you just try?
[Options: 
1 Brainstorming Ideas
2 Choosing / Creating 3D Virtual Object(s)
3 Object Placement
4 Object Selection & Modification
5 Object Selection & Remove Object
]

2. Which version did you use? (See app for the feature version)
[Options: A, B, C]

[If "1 Brainstorming Ideas" was chosen:]
3. In that brainstorming session, how do you feel about the end result, creatively speaking?
[Options: Not at all creative to Very creative on a 5-point scale]

[If "2 Choosing / Creating 3D Virtual Object(s)" was chosen:]
4. When you chose or created an object, how did you feel about the end result, creatively speaking?
[Options: Not at all creative to Very creative on a 5-point scale]

[UMUX-LITE: Both questions]

[NASA-TLX: The Mental Demand, Performance, Effort, and Frustration questions]

\end{lstlisting}

\begin{lstlisting}[caption={The second comparison task questionnaire, which participants completed after they finished all of the trials. This questionnaire compared the systems overall (as opposed to each feature).},label={lst:all-feature-comparison-survey},frame=single]
1. Which system did you prefer? (Please choose one, if possible)
[Options:
System A: Manual (e.g. tap to place)
System B: AI system with multiple responses/options (e.g. select one of the AI's object placements)
System C: AI system with single response/option (e.g. AI system places an object)
]

2. Why did you prefer the above system? (Please write at least 1-2 full sentences.)
[Long answer]

\end{lstlisting}

\begin{lstlisting}[caption={The third comparison task questionnaire, which asked participants to brainstorm about new features that had not yet been added to the app.},label={lst:new-feature-comparison-survey},frame=single]
1. Imagine you were adding music to your AR experience. How would you prefer to do this? (Please choose one, if possible)
[Options:
System A: Manually scroll through a list of music files and choose one.
System B: Ask the AI for music (e.g. "Can you add upbeat music?"), and it gives you 3 options.
System C: Ask the AI for music (e.g. "Can you add upbeat music?"), and it adds that song.
Other - Write in
]

2. Why did you prefer the above? Please write 1-2 full sentences.

3. Imagine you were adding sound effects to your AR experience. How would you prefer to do this? (Please choose one, if possible)
[Options:
System A: Choose an object (e.g. a dog), then tap "add sound effects", then manually scroll through a list of sound effect files, and choose one.
System B: Choose an object (e.g. a dog), then ask the AI for a sound effect (e.g. "Can you add a barking sound effect?"), and it gives you 3 options.
System C: Tell the AI to generate a sound effect with respect to an object (e.g. "Can you add a barking sound effect to this dog?"), and it adds that sound effect.
Other - Write in
]

4. Why did you prefer the above? Please write 1-2 full sentences.

5. Imagine you were animating objects in your AR experience. How would you prefer to do this? (Please choose one, if possible)
[Options:
System A: Choose an object, then tap "animate". Tap a start location and end location for the object's journey. Choose which type of animation (e.g. bounce, walk, run, etc.).
System B: Choose an object, then ask the AI system to generate an animation (e.g. "Make it walk to the dog bowl"), and it gives you 3 options proposing a path and walk style.
System C: Tell the AI system to generate an animation (e.g. "Make it walk to the dog bowl"), and it generates that animation.
Other - Write in
]

6. Why did you prefer the above? Please write 1-2 full sentences.

7. Imagine you were preparing an object to react to an event in your AR experience. 
***An event*** can be triggered by a person pressing a virtual button, standing near a virtual object, making a noise or saying a trigger word, etc.
***A reaction*** to an event could be an object animating or changing colour.
How would you prefer to prepare an object to react to an event? (Please choose one, if possible)
[Options:
System A: Choose the reacting object, then tap "events". Choose an event from a dropdown list. Choose how the object should react from a list.
System B: Choose the reacting object, then tell the AI system what event should trigger what corresponding reaction. The AI gives you 3 options proposing various reactions.
System C: Tell the AI system that an event should trigger a reaction. The AI provides that reaction to that event.
Other - Write in
]

8. Why did you prefer the above? Please write 1-2 full sentences.

9. Imagine you were pinning a virtual object to a real one in your AR experience. E.g. pinning a virtual hat to a statue.
How would you prefer to do this? (Please choose one, if possible)
[Option:
System A: Choose the real object, e.g. the head of a statue. Tap "pin" and then choose a virtual object from a list, e.g. a hat. Adjust the virtual object's location/scale/orientation to fit the real object.
System B: Choose the real object, then tell the AI system what virtual object to pin. The AI gives you 3 options proposing various objects and locations/scales/orientations.
System C: Tell the AI to pin a virtual object to a real object, and it pins it.
Other - Write in
]

10. Why did you prefer the above? Please write 1-2 full sentences.
[Long answer]

11. Any other comments?
[Long answer]

\end{lstlisting}

\subsection{Part 2: Free-Form Authoring Task Questionnaire}
After participants completed the free-form authoring task in Part 2 of the study, they completed the Creativity Support Index (CSI) questionnaire \cite{Cherry2014}. This can be found in \citeauthor{Cherry2014}'s paper \cite{Cherry2014}. Note that we did not include the Collaboration factor, as our system is not collaborative.

\subsection{Part 3: Semi-Structured Interview Questions}
Listing~\ref{lst:interview-questions} contains the main interview questions we asked participants. Since it was a semi-structured interview, we additionally asked follow-up questions about certain responses, \eg, for clarity.

\begin{lstlisting}[caption={The main interview questions we asked participants in Part 3 of the study.},label={lst:interview-questions},frame=single]
~ Questions about the tool ~
1. What did you build and who did you build it for?

2. What features did you use to build it? Why?

3. Did any of the features you used while building stand out to you? Why?
  a. What was your favourite feature? Why?
  b. What was your least favourite feature? Why?

4. What were some of the pros/cons to using the AI agent vs manual process?

5. Was there anything that you found particularly enjoyable or rewarding?

6. Was there anything that you found particularly stressful?
  a. (E.g. social stressors)

7. Imagine that your job was to create AR experiences (e.g. for the audience you chose in Q1). Would you feel confident using this as an initial brainstorming/prototyping tool? Why?

~ Brainstorming questions ~
8. Imagine you could make this tool infinitely better. How might you improve it? What two features might you add?

9. Imagine you had this "better tool" and much more time, and you could build an AR experience in any location. What AR experience might you build and where?

~ Final question ~
10. Any other thoughts or comments?

\end{lstlisting}

\bibliographystyle{ACM-Reference-Format}
\bibliography{paper}